\documentclass{pasj00}
\usepackage{times}
\usepackage[T1]{fontenc}

\begin{document}
\SetRunningHead{Y. Doi et.al.}{The {\it AKARI} Far-Infrared All-Sky Survey Maps}
\Received{}%{yyyy/mm/dd}
\Accepted{}%{yyyy/mm/dd}

\title{The {\it AKARI} Far-Infrared All-Sky Survey Maps}

%%% begin:list of authors
% Do NOT capitalize all letters in "textsc".
\author{
  Yasuo \textsc{Doi},\altaffilmark{1}
  Satoshi \textsc{Takita},\altaffilmark{2}
  Takafumi \textsc{Ootsubo},\altaffilmark{1}
  Ko \textsc{Arimatsu},\altaffilmark{2}
  Masahiro \textsc{Tanaka},\altaffilmark{3}
  Yoshimi \textsc{Kitamura},\altaffilmark{2}
  Mitsunobu \textsc{Kawada},\altaffilmark{2}
  Shuji \textsc{Matsuura},\altaffilmark{2}
  Takao \textsc{Nakagawa},\altaffilmark{2}
  Takahiro \textsc{Morishima},\altaffilmark{4}
  Makoto \textsc{Hattori},\altaffilmark{4}
  Shinya \textsc{Komugi},\altaffilmark{2}\thanks{Present address: Division of Liberal Arts, Kogakuin University, 2665-1 Nakano-machi, Hachioji, Tokyo 192-0015}
  Glenn J. \textsc{White},\altaffilmark{5,6}
  Norio \textsc{Ikeda},\altaffilmark{2}
  Daisuke \textsc{Kato},\altaffilmark{2}
  Yuji \textsc{Chinone},\altaffilmark{4}\thanks{Present address: High Energy Accelerator Research Organization (KEK), Tsukuba, Ibaraki 305-0801}
  Mireya \textsc{Etxaluze},\altaffilmark{5,6}
  and
  Elysandra \textsc{Figueredo}\altaffilmark{5,6}\thanks{Present
    address: Department of Astronomy, IAG, Universidade de S\~{a}o Paulo, 05508-090 S\~{a}o Paulo, SP, Brazil}
}
\altaffiltext{1}{Department of Earth Science and Astronomy, University of Tokyo, Komaba 3-8-1, Meguro, Tokyo 153-8902}
\email{doi@ea.c.u-tokyo.ac.jp}
\altaffiltext{2}{Institute of Space and Astronautical Science, Japan Aerospace Exploration Agency, 3-1-1 Yoshinodai, Sagamihara, Kanagawa 252-5210}
\altaffiltext{3}{Center for Computational Sciences, University of
  Tsukuba, 1-1-1,Tennodai Tsukuba-city, Ibaraki 305-8577}
\altaffiltext{4}{Astronomical Institute, Tohoku University, Aoba-ku, Sendai 980-77}
\altaffiltext{5}{Department of Physics \& Astronomy, The Open University, Milton Keynes, MK7 6BJ, United Kingdom}
\altaffiltext{6}{Space Science and Technology Dept., The Rutherford Appleton Laboratory, Didcot, OX11 0QX, United Kingdom}
%%% end:list of authors

%%% Please use the following style in case that sorting by
%%% affilation is impossible.
%
% \author{%
%   D-Firstname \textsc{D-Familyname}\altaffilmark{1}
%   E-Firstname \textsc{E-Familyname}\altaffilmark{1,2}
%   and
%   F-Firstname \textsc{F-Familyname}\altaffilmark{2}}
% \altaffiltext{1}{Address of Institute}
% \email{ddddd@xxx.xxx.xx.xx}
% \email{eeeee@xxx.xxx.xx.xx}
% \altaffiltext{2}{Address of Institute}

%% `\KeyWords{}' always has to be placed before `\maketitle'.

\maketitle

\begin{abstract} %%% Abstract to run on from here.

We present a far-infrared all-sky atlas from a sensitive all-sky
survey using the Japanese {\it AKARI} satellite.
The survey covers $> 99$\% of the sky in four photometric bands centred
at 65 \micron, 90 \micron, 140 \micron, and 160 \micron\ with spatial
resolutions ranging from 1 to 1.5 arcmin.
These data provide crucial information for the investigation and characterisation of the
properties of dusty material in the Interstellar Medium (ISM), since
significant portion of its energy is emitted
between $\sim50$ and 200 \micron.
The large-scale distribution of interstellar clouds, their thermal dust temperatures and column densities, can be investigated with the improved spatial
resolution compared to earlier all-sky survey observations.
In addition to the point source distribution, the large-scale distribution of ISM cirrus emission, and its filamentary structure, are well traced.
We have made the first public release of the full-sky data to provide a legacy data set for use by the astronomical community.

\end{abstract}

%%% MAIN BODY OF TEXT GOES HERE. CONSULT "INSTRUCTIONS FOR AUTHORS USING
%%% LATEX2E MARKUP", SECTIONS 2.3-2.6 FOR HELP WITH EQUATIONS, FIGURES,
%%% AND TABLES.

%\section{}   %%% Top level section head (remove "%" symbol)
%\subsection{}   %%% Second level section head (remove "%" symbol)
%\subsubsection{}   %%% Lowest level section head (remove "%" symbol)
%\section*{}    %%% Unnumbered top level section head (remove "%" symbol)
%\subsection*{}   %%% Unnumbered second level section head (remove "%" symbol)

\section{Introduction}\label{sec:intro}

Infrared continuum emission is ubiquitous across the sky, and is attributed to the thermal emission from interstellar dust particles.
These interstellar dust particles are heated by incident stellar radiation field of UV, optical, and near-infrared wavelengths, which together have an energy peak at around 1 \micron\ (interstellar radiation field: ISRF; \cite{Mathis83}).
The heated dust radiates its thermal energy at longer wavelengths in
the infrared to mm wavelength range.
The spectral energy distribution (SED) of the dust continuum emission
has been observed from various astronomical
objects including the diffuse interstellar medium (ISM), star-formation regions and galaxies.
The observed SEDs show their peak at far-infrared (30 -- 300 \micron;
FIR) wavelengths, with approximately two thirds of the energy being radiated at $\lambda \geq 50$ \micron\ (\cite{Draine03}; also see \cite{Compiegne11}).
It is thus important to measure the total FIR continuum emission
energy as it is a good tracer of total stellar radiation energy, which
is dominated by the radiation from young OB-stars, and thus a good
indicator of the star-formation activity (\cite{Kennicutt98}).

The first all-sky survey of infrared continuum between 12 and 100
\micron\ was pioneered by the IRAS satellite (\cite{Neugebauer84}).
It carried out a photometric survey with two FIR photometric bands
centred at 60 \micron\ and 100 \micron\, achieving a spatial
resolution of $\sim 4'$.
The IRAS observation was designed to detect point sources, but not designed to make
absolute photometry (\cite{Beichman88}),
leaving the surface brightness of diffuse emission with spatial scales larger than $\sim 5'$ to be only relatively measured.
It has been possible, however, to create large-area sky maps (the IRAS Sky Survey Atlas: ISSA; \cite{Wheelock1994}), by filtering out detector sensitivity drifts.
An improvement to the efficacy of the IRAS diffuse emission data was
made by combining absolute photometry data with lower spatial
resolution from COBE/DIRBE observation (\cite{Boggess92}; \cite{Hauser98}),
and by using an improved image destriping technique (Improved Reprocessing of the IRAS Survey: IRIS; \cite{IRIS}).
An all-sky survey at sub-millimetre wavelengths from 350 \micron\ to 10 mm were made by the Planck satellite with an angular resolution from 5$'$ to 33$'$ (\cite{2013arXiv1303.5062P}).

The observed FIR dust continuum SEDs are reasonably well fitted by modified black-body
spectra having spectral indices $\beta \sim ∼ $1.5 -- 2, with
temperatures $\sim $15 -- 20 K
(\cite{Boulanger96}; \cite{Lagache98}; \cite{Draine03}; \cite{Roy10};
\cite{Bernard11}; \cite{2014A&A...571A..11P}; \cite{2014arXiv1410.7523M}).
Dust particles with relatively larger sizes (BG: big grains; $a \geq
0.01$ \micron\ with a typical size of $a \sim 0.1$ \micron) are considered to be the main source of this FIR emission (\cite{Desert90}; \cite{Mathis90}; \cite{Draine07}; \cite{Compiegne11}).
The BGs are in thermal equilibrium with the ambient ISRF
and thus their temperatures ($T_{\rm BG}$) trace the local ISRF intensity ($I_{\rm ISRF}$; \cite{Compiegne10}; \cite{Bernard10}).
Measuring the SED of the BG emitters both above, and below its peak,
is therefore important to determining $T_{\rm BG}$.
For the diffuse ISM, we can assume a uniform $T_{\rm BG}$ along the
line of sight and thus can convert the observed $T_{\rm BG}$ to $I_{\rm ISRF}$.

The opacity of the BGs ($\tau_{\rm BG}$) of the diffuse clouds can be
evaluated from the observed FIR intensity and $T_{\rm BG}$.
Assuming that dust and gas are well mixed in interstellar space, $\tau_{\rm BG}$ is a good tracer of the total gas column density including all the interstellar components: atomic, molecular and ionized gas (e.g. \cite{Boulanger88}; \cite{Joncas92}; \cite{Boulanger96}; \cite{Bernard99}; \cite{Lagache00}; \cite{mamd07}; \cite{Bernard11}).
It is also important to use estimates of $\tau_{\rm BG}$ as a proxy to infer the foreground to the cosmic microwave background (CMB) emission that is observed in
longer wavelengths (\cite{Schlegel98}; \cite{IRIS};
\cite{Compiegne11}; \cite{2013arXiv1303.5062P};
\cite{2014A&A...571A..11P}; \cite{2014arXiv1410.7523M}).

Since the IRAS observations cover only the wavelengths at, or
shorter than 100 \micron,
it is important to note that emission from stochastically heated
smaller dust grains becomes significant at shorter wavelengths.
\citet{Compiegne10} estimated the contribution of smaller dust
emission to their observations with the PACS instrument
(\cite{Poglitsch10}) on-board the Herschel satellite
(\cite{Pilbratt10}) based on their adopted dust model (\cite{Compiegne11}).
They estimated the contribution to be up to $\sim 50\%$ in the 70
\micron\ band, $\sim 17\%$ in 100 \micron\ band, and up to $\sim 7\%$
in the 160 \micron\ band, respectively and
concluded that photometric observations at $\leq 70$ \micron\ should be modelled by taking into account the significant contribution of emission from stochastically heated grains.
To make accurate $T_{\rm BG}$ determination without suffering from the excess emission from
stochastically heated smaller dust grains, the IRAS data
thus need to be combined with longer
wavelength COBE/DIRBE or Planck data (\cite{Schlegel98};
\cite{2014A&A...571A..11P}; \cite{2014arXiv1410.7523M}) or to be fit with a modelled SED of the dust continuum
(\cite{2014arXiv1409.2495P}).
However, COBE/DIRBE data have limited spatial resolution of a 0.7$^{\circ}$-square
field-of-view (\cite{Hauser98}).
Planck data have a spatial resolution that is comparable with that of
IRAS, but the data do not cover the peak of the dust SED.
Observations cover the peak wavelength of the BG thermal emission with
higher spatial resolutions are therefore required.

Ubiquitous distribution of diffuse infrared emission (cirrus emission)
was first discovered by IRAS observations (\cite{Low84}).
Satellite and balloon-observations have revealed that the cirrus emission has no characteristic spatial scales, and is well represented by Gaussian random fields with a power-law spectrum $\propto k^{-3.0}$
(\cite{Gautier92}; \authorcite{Kiss01} \yearcite{Kiss01}, \yearcite{Kiss03}; \authorcite{mamd02} \yearcite{mamd02}, \yearcite{mamd07}; \cite{Jeong05}; \cite{Roy10}), down to sub-arcminute scales (\cite{mamd10}; \cite{Martin10}).
Recently, observations with Herschel with a
high-spatial resolution ($12\arcsec$ -- $18\arcsec$ at 70, 100, 160 and 250 \micron)
resolved cirrus spatial structures in nearby star-formation regions,
showing that they were dominated by filamentary structures.
These filament structures have shown typical widths of $\sim 0.1$ pc, which is
common for all the filaments of those observed in the nearby Gould
Belt clouds (cf. \cite{2011A&A...529L...6A}, also see
\cite{2013A&A...553A.119A} and a comprehensive review by \cite{2013arXiv1309.7762A}).
Prestellar cores are observed to be concentrated on the filaments
(\cite{2010A&A...518L.106K}; \cite{2010A&A...518L.102A}), suggesting
that the filamentary structure plays a prominent role in the
star-formation process (\cite{2014prpl.conf...27A}).
To reveal the mechanism of the very first stages of the star-formation
process, whose spatial scale changes from that of giant molecular
clouds ($\geq 100$ pc) to the pre-stellar cores ($\leq 0.1$ pc), a wider
field survey with a high spatial resolution is required so that we can
investigate the distribution of the cirrus emission across a broad
range of spatial scales, study its nature in more detail, as to remove
its contribution as a foreground to the CMB.

For these reasons described above, we have made a new all-sky survey with an infrared astronomical satellite {\it AKARI} (\cite{Murakami07}).
The sky coverage of this survey was 99\%, with 97\% of the sky covered with multiple scans.
The observed wavelength coverage spans 50 -- 180 \micron\ continuously
with four photometric bands, centred at 65 \micron, 90 \micron, 140
\micron, and 160 \micron\ having spatial resolutions of $1'$ -- $1'\negthinspace.5$.
The detection limit of the four bands reaches 2.5 -- 16 [MJy sr$^{-1}$] with relative accuracy of $<20$\%.

In this paper, we describe the details of the observation and the
data analysis procedure.
We also discuss the suitability of the data to measure the SED and
estimate spectral peak of the dust emission (so as to investigate the
total FIR emission energy), and to determine the spatial distribution
of the interstellar matter and star-formation activity at a high spatial resolution.

In $\S$\ref{sec:observation}, we describe the observation by {\it AKARI} with our FIR instrument.
The details of the data analysis are given in $\S$\ref{sec:analysis},
including subtraction of foreground Zodiacal emission, correction
scheme of transient response of FIR detectors, and image destriping.
Characteristics of the produced images and their quality are described
in $\S$\ref{sec:results}.
In $\S$\ref{sec:discussion},  we describe the capability of the {\it
  AKARI} image to make the detailed evaluation of the spatial
distribution and the SED of the interstellar dust emission.
The caveats that we have about remaining artefacts in the images, and our future plans to mitigate the effects of these are described in $\S$\ref{sec:caveats}.
Information of the data release is given in $\S$\ref{sec:datarelease}.
In $\S$\ref{sec:conclusion}, we summarise the results.

\section{Observation}\label{sec:observation}

We performed an all-sky survey observation with the {\it AKARI} satellite, a dedicated satellite for infrared astronomical observations \citep{Murakami07}.
Its telescope mirror had a diameter of $\phi = 685$ mm
and was capable of observing across the 2 -- 180 \micron\ near to
far-infrared spectral regions, with two focal-plane instruments (FPIs): the InfraRed Camera (IRC: \cite{Onaka07}) and the Far-Infrared Surveyor (FIS: \cite{Kawada07}).
Both of the FPIs, as well as the telescope, were cooled down to 6 K
with liquid Helium cryogen and mechanical double-stage Stirling coolers as a support
\citep{Nakagawa07}, reducing the instrumental thermal emission.
The satellite was launched in February 2006 and the all-sky survey was performed during the period April 2006 -- August 2007, which was the cold operational phase of the satellite with liquid Helium cryogen.

The satellite was launched into a sun-synchronous polar orbit with an
altitude of 700 km, an inclination angle of 98\degree\negthinspace.2,
and an orbital period of 98 minutes, so that the
satellite revolved along the day-night boundary of the earth.
During the survey observations, the pointing direction of the telescope was kept orthogonal to the
sun-earth direction and was kept away from the centre of the earth.
Consequently, the radiative heat input from the sun to the satellite was
kept constant and that from the earth was minimised (see Figure 4 of
\cite{Murakami07}).

With this orbital configuration, the {\it AKARI} telescope continuously
scanned the sky along the great circle with a constant scan speed of
3\arcmin\negthinspace.6 sec$^{-1}$.
Due to the yearly revolution of the earth, the scan direction is
shifted $\sim 4$\arcmin~per satellite revolution in a longitudinal direction or in the cross-scan direction.
Consequently it was possible to survey the whole sky during each six month period of continuous observation.

Dedicated pointed observations of specific astronomical objects were
interspersed with survey observations (\cite{Murakami07}).
The survey observation was halted for 30 minutes during a pointed
observation to allow for 10 minutes of integration time as well as the
satellite's attitude manoeuvre before and after the pointed observation.
The regions that had been left un-surveyed due to pointed observations
were re-scanned at a later time during the cold operational phase.

\begin{figure}[ht]
  \begin{center}
    \FigureFile(80mm,101.6mm){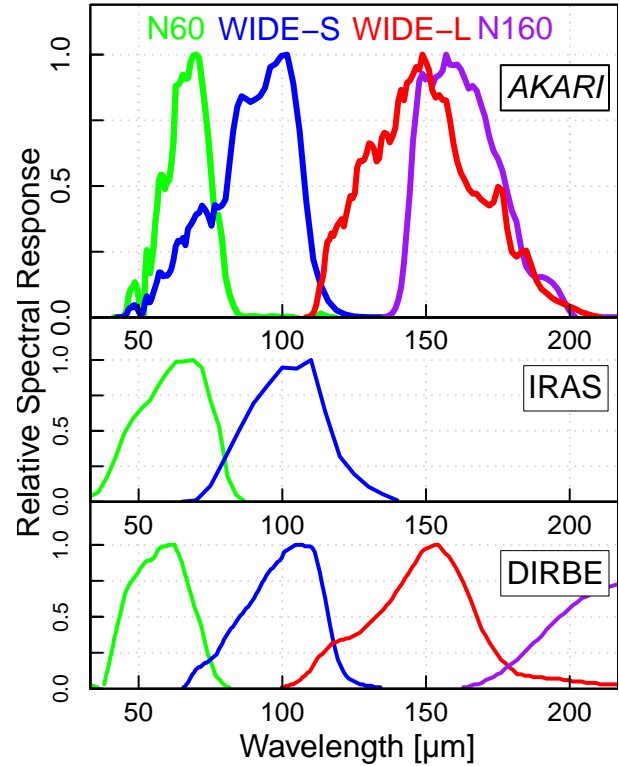}
  \end{center}
  \caption{Spectral responses of the four {\it AKARI} FIR bands
    centred at 65 \micron\ (N60), 90 \micron\ (WIDE-S), 140 \micron\ (WIDE-L),
    and 160 \micron\ (N160). Note that {\it AKARI} has continuous
    wavebands covering 50 -- 180 \micron\ so that we can make precise
    evaluation of the total FIR intensity from the in-band flux of the {\it AKARI}
    observations. Spectral responses of IRAS and COBE/DIRBE
    are also shown for comparison.}\label{fig:spectra}
\end{figure}

The all-sky survey observations at far-IR wavelengths were performed by
using the FIS instrument, which was dedicated for photometric scan and
spectroscopic observations across the 50 -- 180 \micron\ spectral region \citep{Kawada07}.
Four photometric bands were used for the all-sky survey.
The spectral responses of the bands are shown in Figure~\ref{fig:spectra}
and the characteristics of the bands are summarised in Table~\ref{tab:spec}.
\begin{table*}[bht]
  \caption{Specification of the {\it AKARI} FIR detectors (\cite{Kawada07}; \cite{Shirahata09}).}
  \label{tab:spec}
  \begin{center}
    \leavevmode
    \footnotesize
    \begin{tabular}[h]{lccccl}
      \hline \\[-5pt]
      Band name & N60 & WIDE-S & WIDE-L & N160\\[+5pt]
      \hline \\[-5pt]
      Centre wavelength & 65 & 90 & 140 & 160 & [\micron]\\
      Wavelength range & 50 -- 80 & 60 -- 110 & 110 -- 180 & 140 -- 180 & [\micron]\\
      Array format & $20\times 2$ & $20\times 3$ & $15\times 3$ & $15\times 2$ & [pixels]\\
      Pixel scale\footnotemark[1] & \multicolumn{2}{c}{$26\arcsec \negthinspace\negthinspace .8 \times 26\arcsec \negthinspace\negthinspace .8$} & \multicolumn{2}{c}{$44\arcsec \negthinspace\negthinspace .2 \times 44\arcsec \negthinspace\negthinspace .2$} & \\
      Pixel pitch\footnotemark[1] & \multicolumn{2}{c}{$29\arcsec \negthinspace\negthinspace .5 \times 29\arcsec \negthinspace\negthinspace .5$} & \multicolumn{2}{c}{$49\arcsec \negthinspace\negthinspace .1 \times 49\arcsec \negthinspace\negthinspace .1$} & \\
      Detector device & \multicolumn{2}{c}{Monolithic Ge:Ga array} & \multicolumn{2}{c}{Stressed Ge:Ga array} & \\
      \hline \\[-8pt]
      \multicolumn{6}{l}{\footnotesize{\footnotemark[1]At the array centre.}}\\
      \end{tabular}
  \end{center}
\end{table*}
Two of four bands had a broader wavelength coverage and continuously
covered the whole waveband range: the WIDE-S band (50 -- 110 \micron,
centred at 90 \micron) and the WIDE-L band (110 -- 180 \micron, centred at 140 \micron).
The other two bands had narrower wavelength coverage
and sampled both the shorter and the longer ends of the wavebands: the N60 band
(50 -- 80 \micron, centred at 65 \micron) and the N160 band (140 -- 180 \micron, centred at 160 \micron).

The pixel scales of the detectors were $26\arcsec\negthinspace\negthinspace .8 \times
26\arcsec\negthinspace\negthinspace .8$ for the
short wavelength bands (N60 and WIDE-S) and $44\arcsec\negthinspace\negthinspace .2 \times
44\arcsec\negthinspace\negthinspace .2$ for the long wavelength bands (WIDE-L and N160; Table~\ref{tab:spec}).
The cross-scan width of the detector arrays were $\sim 8\arcmin$ for N60 and
WIDE-S and $\sim 12\arcmin$ for WIDE-L and N160 (see Figure 3 of \cite{Kawada07}).
These array widths corresponded to two or three times the shift of the scan direction per satellite revolution ($\sim 4\arcmin$, see above).
With a continuous survey observation, as a result, regions close to the ecliptic plane were surveyed at least twice with the N60 and WIDE-S bands and three times with the WIDE-L and N160 bands. Regions at higher ecliptic latitudes had increasingly greater exposure times and numbers of confirmatory scans.

The detector signals were read out by Capacitive Trans-Impedance
Amplifiers (CTIAs) that were specially developed for this satellite mission \citep{Nagata04}.
The CTIA is an integrating amplifier so that the detector signal was
read out as an integration ramp, with a slope that is proportional to the detector flux.
The signal level was sampled at 25.28Hz (N60, WIDE-S) and 16.86Hz (WIDE-L, N160), which corresponded to about three samples in a pixel crossing time of an astronomical source \citep{Kawada07}.

The integrated electrical charge was reset periodically so that
the output voltage remained well within the dynamic range of the CTIA.
We applied regular reset intervals of either 2 seconds, 1 second, or after every sampling
(correlated double sampling: CDS) for the all-sky survey observations
depending upon the sky brightness, with reference to the sky brightness at 140 \micron\ observed by CORBE/DIRBE from Annual Average Map \citep{Hauser98}.
A reset interval of 2-sec was normally applied, although it was shortened to 1-sec for the brighter celestial regions with $> 60$ [MJy sr$^{-1}$].
The CDS mode was applied for sky with $> 210$ [MJy sr$^{-1}$], which mainly corresponds to the inner Galactic plane.

An additional calibration sequence was periodically carried out during the survey observations to calibrate the absolute and relative sensitivities of the various detector channels.
The FIS was equipped with a cold shutter to allow absolute sensitivity calibration \citep{Kawada07}.
The shutter was closed for 1 minute after every 150 minutes of
observations, or after $\sim 1.5$ orbits of the satellite, so as to estimate the dark-sky condition, and to measure the absolute zero level of the detector signal.
A constant illumination calibration flash was also applied for 30 seconds while the shutter was closed so that we could calibrate the responsivity of the detector channels in addition to the dark measurement.
This calibration sequence was performed when the regions near the north or the south ecliptic pole were surveyed as the regions were repeatedly surveyed with many survey scans.
The observed regions that were affected by this calibration
sequence were intentionally scattered in the pole regions so that we could maximise the on-sky observing time.

In addition to the absolute calibration sequences mentioned above, a
calibration light was flashed every minute during the survey
observations to calibrate sensitivity drifts whose time variations
were shorter than 150 minutes.
The flash simulated the profile of a point source crossing the FOV of each detector pixel.

\begin{figure*}
   \begin{center}
      \FigureFile(160mm,114.3mm){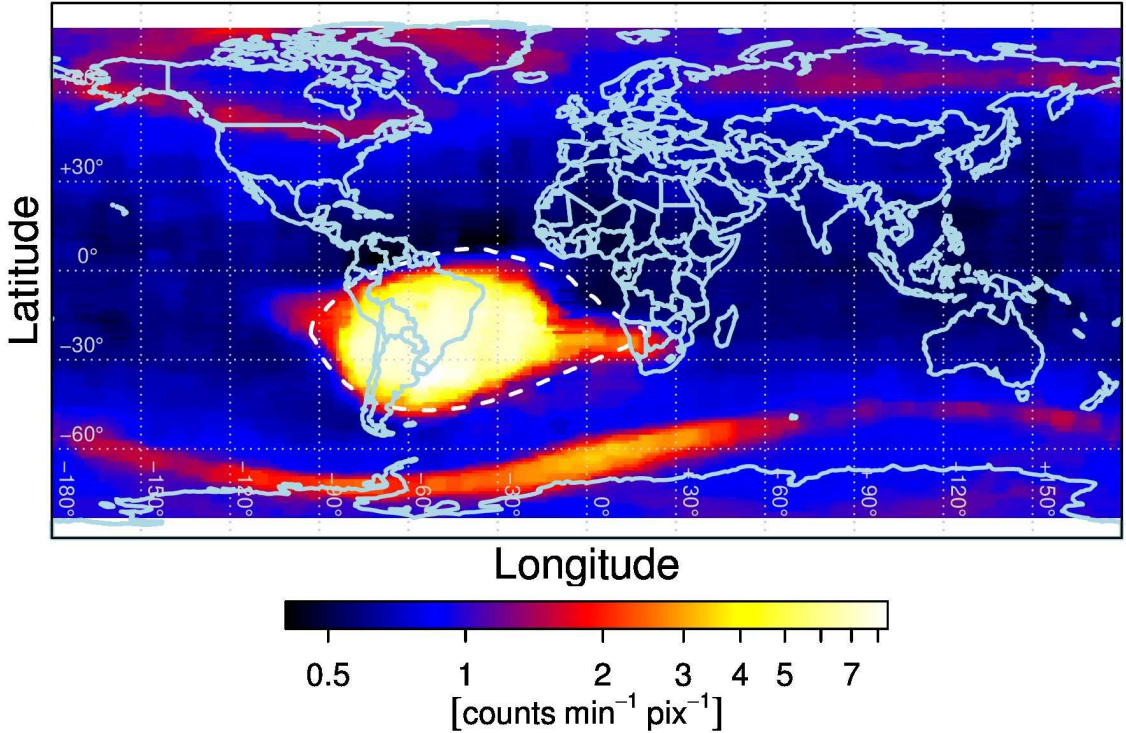}
   \end{center}
  \caption{The spatial distribution of the high-energy particle hit rate evaluated from in-flight data of the FIS detector signal.
The dashed line indicates the region where we stopped the FIS observation during the satellite passages.}
\label{fig:SAA}
\end{figure*}

High-energy particles hitting the detector cause glitches or sudden jumps in the detector signals \citep{Suzuki08}, which must be restored or removed from analyses to eliminate false detection of astronomical objects.
However, the high frequency of high-energy particle hits prevents continuous observations, and may even cause significant drift of the detector responsivities.
A region above the earth's surface over the South Atlantic is known to have anomalous geomagnetic field which leads to a higher density of solar protons (the South Atlantic Anomaly: SAA).
We characterized this region from in-flight data of particle hit rates as shown in Figure~\ref{fig:SAA}.
We paused the survey observation during satellite passages through this region.

During the SAA passages, the response of each detector pixel was
significantly changed due to high frequency hitting of high-energy
particles, which led to excitation of charged particles in the detector material \citep{Suzuki08}.
To restore the detector responsivity, we applied a high electrical voltage (on the order of 1 V) for 30 seconds to each detector pixel to
flush out the charged particles (bias-boosting).
We performed a bias-boost operation after every SAA passage waiting 1 minutes after exit from the SAA region, re-starting the survey observations two minutes after the bias-boost procedure.

The scattered light of bright objects from the telescope structures at the periphery of the telescope's field of view can also affect the survey data, and emission from the moon was found to be the dominant contributor to this contamination.
\begin{figure}[ht]
  \begin{center}
      \FigureFile(80mm,80mm){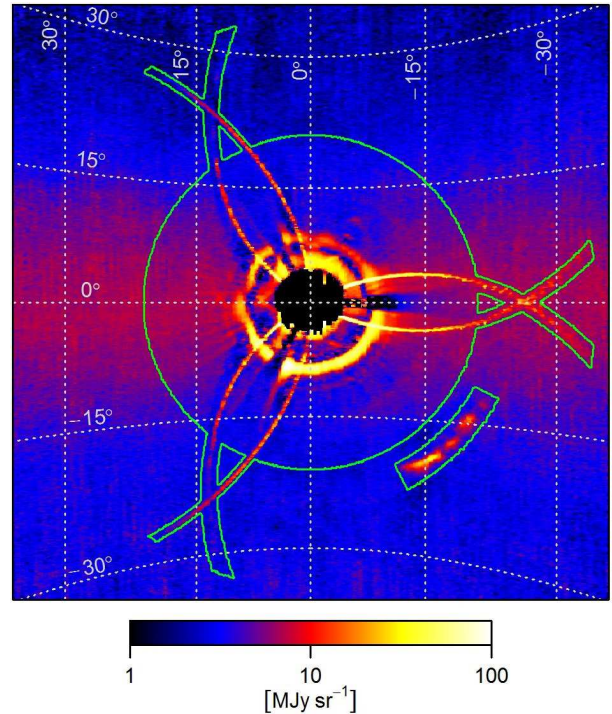}
  \end{center}
  \caption{Scattered light around the moon observed in the Wide-L
    band. The moon position is at the centre of the figure. We closed
    the shutter near the moon $<5$\degree\ and thus no data are
    available at the region around the moon. The broad diffuse
    horizontal feature in the middle of the figure is the Zodiacal
    dust band component, which has not been removed
from the data (see $\S$\ref{sec:zodi}). The green lines indicate the regions which are contaminated by the scattered light. The data observed in these regions were eliminated from the image processing.}
\label{fig:moon_mask}
\end{figure}
Thus we measured the scattered light pattern as shown in the
Figure~\ref{fig:moon_mask} and eliminated the contaminated data from
the image processing ($\S$\ref{sec:analysis}). The eliminated region is indicated in
Figure~\ref{fig:moon_mask} by the green lines.
The scattered light from bright planets (Jupiter and Saturn) also needs to be considered, but we have not currently applied a correction for these signals, and will defer this for a later reprocessing of the image data ($\S$\ref{sec:cavmoving}).

Interruption of the survey observations due to SAA passages, interference from the moon, and pointed observations left gaps in the sky coverage in the first six-month's survey coverage.
These gaps were filled in at later times using {\it AKARI}'s attitude control, which allowed cross-scan offsets $< \pm 1^{\circ}$.

\begin{figure}[ht]
  \begin{center}
    \FigureFile(80mm,160mm){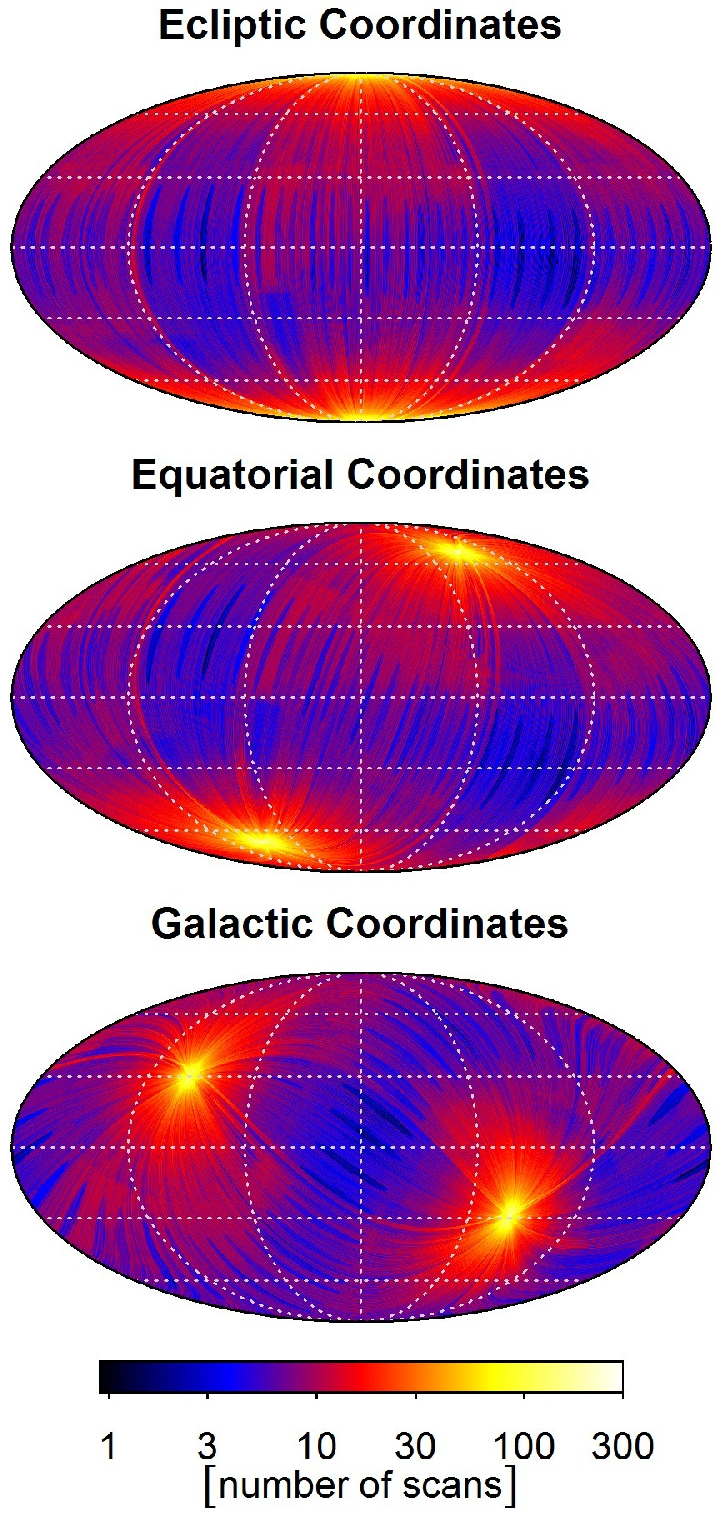}
  \end{center}
  \caption{Sky coverage of the {\it AKARI} all-sky survey. Spatial
    scan numbers are displayed in different celestial coordinates.}
\label{fig:cov}
\end{figure}

A summary of the survey coverage and completeness is shown in Figure~\ref{fig:cov} and Table~\ref{tab:cov}.
\begin{table*}[tbh]
  \caption{Scan coverage of the survey observations.}
  \label{tab:cov}
  \begin{center}
    \leavevmode
    \footnotesize
    \begin{tabular}[h]{lccrrr}
      \hline \\[-5pt]
      Scan coverage & Covered region & Multiply covered region & $\geq5$ times & $\geq10$ times & Not covered\\[+5pt]
      \hline \\[-5pt]
      N60\dotfill & 99.1\% & 96.9\% & 60.3\% & 13.5\% & 0.9\% \\
      Wide-S\dotfill & 99.1\% & 97.0\% & 61.0\% & 13.8\% & 0.9\% \\
      Wide-L\dotfill & 99.5\% & 98.4\% & 78.7\% & 25.9\% & 0.5\% \\
      N160\dotfill & 99.5\% & 98.4\% & 76.8\% & 24.1\% & 0.5\% \\[+5pt]
      \hline\\[-8pt]
      4 bands\dotfill & 99.1\% & 96.8\% & 60.0\% & 13.2\% & 0.9\% \\[+5pt]
      \hline \\[-8pt]
      \end{tabular}
  \end{center}
\end{table*}
After about 17 months of the survey period (the cold operational
phase), 97 \% of the sky was multiply surveyed in the four photometric bands.

\section{Data Analysis}\label{sec:analysis}

The data obtained during the observations were pre-processed using the {\it AKARI} pipeline tool originally optimised for point source extraction \citep{Yamamura09}.
This included corrections for the linearity of the CTIA amplifiers and
sensitivity drifts of the detectors referenced against the calibration
signals (\S\ref{sec:observation}), rejection of anomalous data due to high-energy particle (glitches), signal saturation, and other instrumental effects as well as dark-current subtraction (\cite{Kawada07}; \cite{Yamamura09}).
The orientation of the detector FOV was determined using  data taken by a NIR star camera \citep{Yamamura09}.

Following this initial pre-processing, a separate pipeline was then used to accurately recover the large-scale spatial structures.
The main properties of the diffuse mapping pipeline are as follows: 1) transient response correction of the detector signal, 2) subtraction of zodiacal emission, 3) image processing, 4) image destriping, and 5) recursive deglitching and flat-fielding to produce the final image.
Processes 1) and 2) are performed on the time-series data and 3) -- 5) are performed on image plane data.
In the following section, we describe the details of these procedures.

\subsection{Time-series data analysis with transient response correction}\label{sec:timeline}

The time-series signal from each detector pixel was processed to recover the true sky flux.
Firstly we eliminated anomalous data that were detected during
pre-process including glitches and data for which the output of the CTIAs became non-linear because of signal saturation.
The data collected during CTIA resets and calibration lamp flashes were also removed (see $\S$\ref{sec:observation} for the amplifier reset and calibration lamp operation).

The data were then converted to astronomical surface brightness [MJy sr$^{-1}$] from the raw output signal values, multiplied by calibration conversion factors (\cite{Takita14}).
Different conversion factors were applied to the data taken in the nominal integration mode, and in the CDS mode ($\S$\ref{sec:observation}) so that these data are treated separately (\cite{Takita14}).

The transient response, due to non-linear behaviour of the detector,
is a major cause of distortion of the detector time-series signal \citep{Kaneda09} and its correction is particularly important for the recovery of high quality diffuse structure \citep{Shirahata09}.
\citet{Kaneda09} have previously discussed the characterisation of
detector slow-response effects and their mitigation, and we take a
practical approach here to correct the all-sky survey data so that
high quality diffuse maps can be recovered with realistic
computational overheads (see \cite{Doi09a} for the details of the
computation scheme).

\begin{figure}[ht]
  \begin{center}
    \FigureFile(80mm,90mm){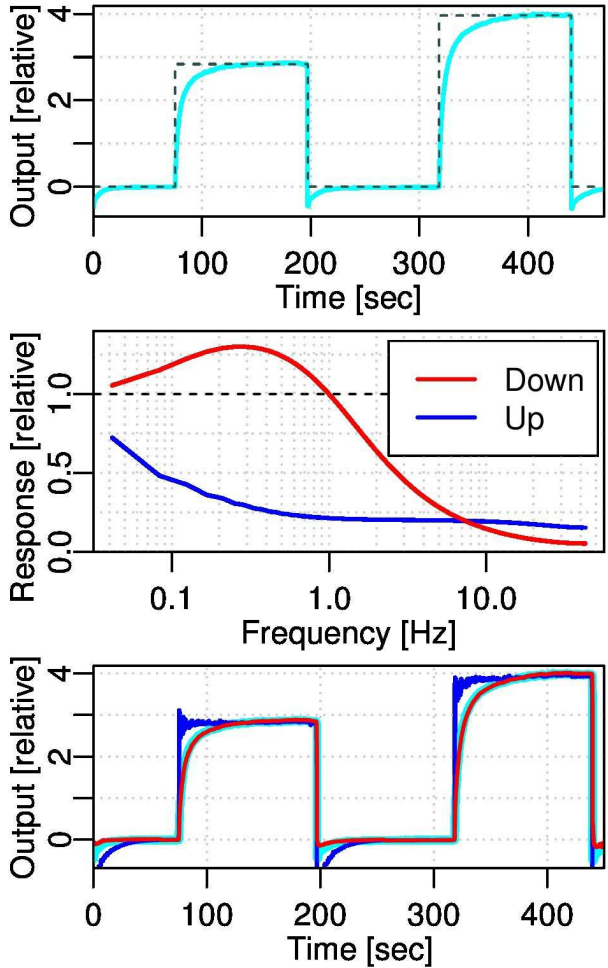}
  \end{center}
  \caption{Top panel -- a step function of the detector signal taken
    in-flight by an internal calibration lamp illumination. The cyan
    line shows the detected signal and the black broken line shows a
    schematic of the lamp illumination pattern. Middle
    panel -- spectral response functions of a detector pixel. The blue
    line shows a spectral response that is estimated from the upward
    step function and the red line shows a spectral response that is
    estimated from the downward step function. Correction functions
    for the transient response of the detector are evaluated from
    these spectral responses (see text). Bottom panel -- signal
    correction by the correction functions evaluated from the above
    spectral responses. The cyan line is the raw detector signal,
    which is the same signal shown in the top panel. The blue line is
    a corrected signal by applying the upward correction function and
    the red line is a corrected signal by applying the downward
    correction function. A reasonable signal profile is reproduced by taking the upper envelope
    of the two corrected signals.
  }
\label{fig:sresponse}
\end{figure}

Figure~\ref{fig:sresponse} top panel shows an example detector
signal that was taken during an in-flight dedicated calibration
sequence.
The cold shutter was closed to achieve the absolute zero level
($\S$\ref{sec:observation}), meanwhile constant
illumination with internal calibration lights was performed with two
different luminosities ($t = 80 - 200$ [sec] and $t = 320 - 440$
[sec]).
The output signal should be rectangular wave (the black broken line in the
figure) if the detector had an
ideal time response, while the observed signal showed significant
distortion due to the transient response.
Upward steps of the observed signal showed slow time response that took
several tens of seconds to reach a stable signal level.
On the other hand, downward steps showed relatively quick response with
significant overshoot at the falling edges.

Since these signals are step response functions to the incident light,
we can evaluate frequency response functions of the detector as the
Fourier transformation of the detector signal
(Figure~\ref{fig:sresponse} middle panel).
The response became low at higher frequencies due to slow-response
effect of the detector.
In addition to that, downward response showed signal amplification at $<1$ [Hz], which
corresponds to the overshoot of the detector signal.

The transient response of the raw detector signal can be corrected by
referring these frequency response functions.
The correction function should have a frequency response that is
inverse of that of the detector, so the correction function is
obtained by deriving reciprocal
of the detector frequency response and then make Fourier inverse
transformation of the reciprocal data.
Since the response function for upward steps and that for downward steps
were significantly different, we derived both upward and downward
transient correction functions and applied both to the same detector
signal (Figure~\ref{fig:sresponse} bottom panel).
A reasonable correction is achieved by taking the larger signal at each
sampling, which is the upper envelope of the two corrected signals.

Because the high-frequency signal component is amplified with the
transient response correction, the high-frequency noise is also
amplified.
To mitigate this side effect,
we suppressed the high-frequency part of the numerical filtering
to filter out the frequency components having higher frequencies than the signal crossing time of the detector FOV.
The residual glitch noises should be eliminated by a deglitching process described below ($\S$\ref{sec:image_process}; $\S$\ref{sec:deglitching}).

\begin{figure}[ht]
\begin{center}
\FigureFile(80mm,48mm){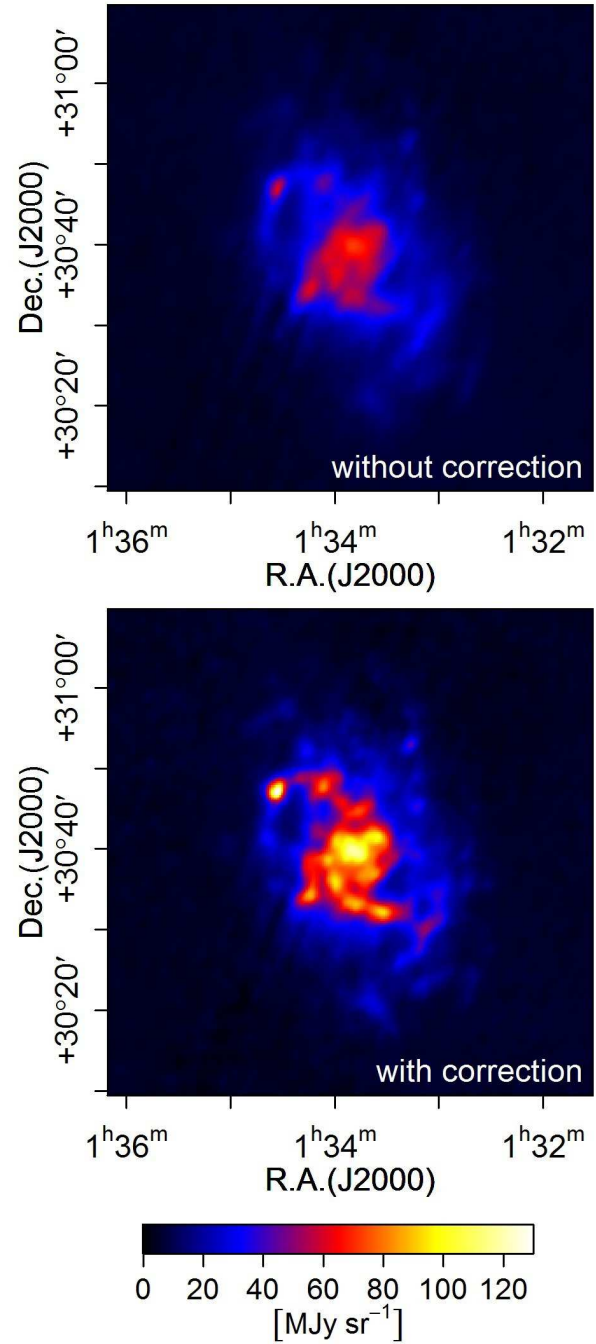}
\end{center}
   \caption{
     An example of the transient response correction that is applied
     to M33 WIDE-L images.
     The top panel shows the image without the correction for the comparison to the image with the correction in the bottom panel.
     An image destriping process ($\S$\ref{sec:destriping}) has been applied for both images.
   }\label{fig:correction}
\vspace*{-3mm}
\end{figure}

\subsection{Subtraction of Zodiacal emission}\label{sec:zodi}

The zodiacal light emission (ZE) is the thermal emission from the interplanetary
dust (IPD) and the dominant diffuse radiation in the mid-infrared (MIR) to FIR
wavelength regions.
Since the dust around 1 AU from the sun has a thermal
equilibrium temperature of $\sim 280$ K, the ZE has a
spectral peak around 10 -- 20 \micron, and it dominates the MIR
brightness in the diffuse sky.
Even in the FIR, the ZE contribution is not completely negligible at
high galactic latitude.
Therefore the ZE component should be segregated from the observed
signal to make astronomical sky maps.

Many efforts have been devoted to describe the structure of the
zodiacal dust cloud.
In particular, number of models have been developed using the infrared satellite data, such as IRAS and
COBE/DIRBE. The ZE model most commonly used to date is the one based
on the DIRBE data ({\it e.g.} \cite{Kelsall98};
\cite{1998ApJ...496....1W}; hereafter the Kelsall model and the Wright
model).
The Kelsall and the Wright models explain the IPD cloud complex with the
following three components:
a smooth cloud, asteroidal dust bands, and a mean motion resonance (MMR)
component.
The MMR component is further divided into a circumsolar ring and a blob
trailing the earth in the Kelsall model.
For each of the three components, they introduced parametrised functions
to describe
the spatial distribution of the dust number density in the heliocentric
coordinates.
\authorcite{2000ApJ...536..550G} (\yearcite{2000ApJ...536..550G}; hereafter the Gorjian model) investigated the DIRBE ZE model
in depth and modified the parameters of
the Wright model.

Although the most part of the ZE structure in {\it AKARI}
FIR images can be well reproduced with the Kelsall or the
Wright/Gorjian models, there are discrepancies at small-scale structures.
In particular, the intensity and the ecliptic latitudes
of the peak positions of the asteroidal dust bands cannot be precisely
reproduced with these models.
\citet{2010A&A...523A..53P} also reported an inconsistency of 20\% for
the intensity of the ring component
between the {\it AKARI} observations and the Kelsall model prediction in MIR.
From these points of view, we only subtracted the smooth cloud component of
the ZE based on the Gorjian model at this stage.
The contribution of the asteroidal dust bands and the MMR component will be
described in a separate paper (\cite{Ootsubo14}), which will provide a ZE model including all components for {\it
AKARI} FIS all sky images.

The Gorjian model evaluates the expected ZE brightness in the DIRBE bands.
Based on this model evaluation, we estimated the SED of the diffuse ZE
component by interpolating the expected ZE brightness in 25, 60, 100,
140, and 240 \micron\ DIRBE bands using a cubic Hermite spline.
Then the expected diffuse ZE brightness in each
{\it AKARI} band was estimated by multiplying this SED by each
{\it AKARI} band (Figure~\ref{fig:spectra}).
The earth's orbital position in the solar system at the time of the
observation and the viewing direction from the satellite were also
considered in the model estimation to take the non-uniform spatial distribution of the IPD cloud in
the solar system into account.

The intensities of dust band and MMR components remaining in the
released {\it AKARI} FIR images are less than about 5 [MJy~sr$^{-1}$] for
all bands.
Detailed values and caveats about the ZE
are given in $\S$\ref{sec:cavzodi}.

\subsection{Image processing from time-series signals}\label{sec:image_process}

The processed time-series data described above were used to produce a preliminary intensity image.
To evaluate the intensity of an individual image pixel, we selected data samples near the pixel centre.
The selection radius was set at three times the half power beam width (\cite{Shirahata09}; \cite{Takita14}).

Outlier data were removed from the samples.
The outliers are mainly due to 1) glitch residuals, 2) noise signal
amplified by the transient response correction, and 3) base-line offsets due to long-term sensitivity variations of the detector.
The outliers 1) and 2) were eliminated by using a sigma clipping method, and
at this level about half of the data samples were eliminated.
As for the outliers 3), a string of data was eliminated by referring whose intensity
distribution function was distinct from other data.
These eliminated data were restored with an additional gain and offset
adjustment in a later stage of the data processing ($\S$\ref{sec:deglitching}).

The residual data were weighted by distance from the pixel centre by
assuming a Gaussian beam profile, whose full width at half maximum was $30\arcsec$
for N60 and WIDE-S, and $50\arcsec$ for WIDE-L and N160, so that the beam
width was comparable to the spatial resolution of the photometric
bands and did not therefore degrade the spatial resolutions of resultant images significantly.
The weighted mean of the residual data was then taken as the preliminary intensity of the image pixel.
The weighted standard deviation, sample number, and spatial scan numbers were also recorded.

\subsection{Image destriping}\label{sec:destriping}

The resultant images show residual spatial stripes due to imperfect
flat-fielding caused by long-term sensitivity variations of the
detector. To eliminate this artificial pattern, we developed a
destriping method based on \citet{IRIS}, who developed their
destriping method to obtain the Improved Reprocessing of the IRAS
Survey (IRIS) map.
The details of our procedure will be described in \citet{Tanaka14}. We describe a brief summary in this paper.

The destriping method by \citet{IRIS} can be summarised as follows:
\begin{enumerate}
\item Image decomposition: An input image is decomposed into three
  components; a large-scale emission map, a small-scale emission map,
  and a point sources map.
\item Stripe cleaning: The small-scale emission map is
  Fourier-transformed to obtain a spatial frequency map.
  Spatial frequency components that are affected by the stripes are
  confined in the spatial frequency map along the radial directions that
  correspond to the stripe directions in the original intensity
  map. The affected spatial frequency components are replaced with
  magnitudes that have averaged values along the azimuthal direction
  (see Figure 1 of \cite{IRIS}).
\item Image restoration: The stripe-cleaned map is inversely
  Fourier-transformed and combined with the other decomposed images (a
  large-scale emission map and a point-sources map).
\end{enumerate}
We find that the stripe cleaning process of the small-scale emission map
described above causes strong artefact in some of our images containing bright molecular
clouds (e.g., Orion clouds), since such regions contain small-scale
and large-amplitude fluctuations of emission.
Such large intensity fluctuations contaminate the derived spatial spectrum and
degrade the outcomes of subsequent destriping processes.
Therefore, we masked the regions that showed large intensity variance in a
small-scale map and eliminated those regions from the stripe cleaning
process, in addition to a large-scale emission map and a point-sources map.
As an indicator of the variance, we calculated the local Root Mean Square
(RMS) of the pixel values on a small-scale map and eliminated the regions whose local RMS was greater than a
threshold value (2$\times$RMS of entire region).
We need to apodise the discontinuity of pixel values at the
boundaries of the eliminated regions to suppress spurious
spectrum patterns in the estimated spatial frequency spectrum.
Thus we multiplied the pixel values in the eliminated regions by
$(1+(s-1)^2)^{-1}$, where $s$ is (local RMS)/threshold, instead of
filling the regions with zero.

\begin{figure}
 \begin{center}
   \FigureFile(80mm,93mm){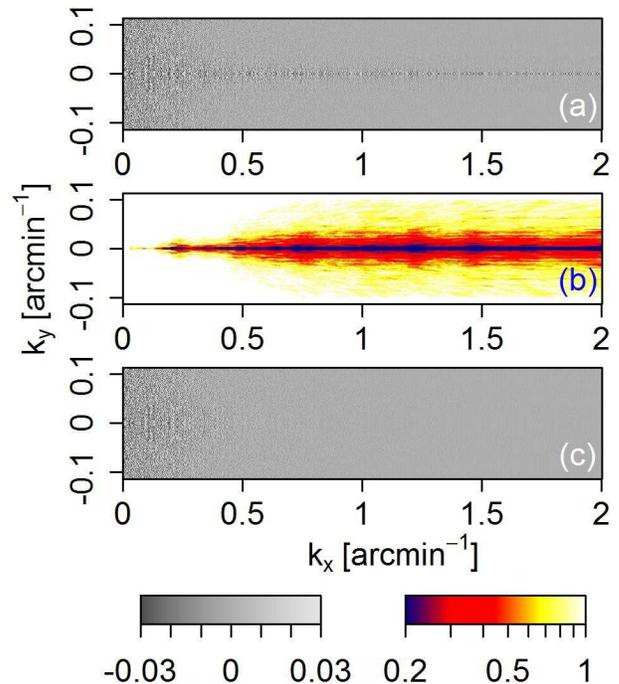}
 \end{center}
\caption{Panel (a) -- A spatial frequency map before destripe, obtained with
  Fourier transformation of an original intensity map containing stripe
  noises. The real part of the complex Fourier moduli is plotted for panels
  (a) and (c).
  Panel (b) -- A map of an attenuation factor, $A(r,\theta)$. See text for
  the definition of $A(r,\theta)$.
  Panel (c) -- A destriped spatial frequency map, obtained by multiplying (a)
  by (b).
  We show details of the maps around the x-axis ($x \geq 0$ [arcmin$^{-1}$], $|y| \leq 0.11$ [arcmin$^{-1}$]), as the excess component caused by the stripe noises is confined along
  the x-axis (see text).
  The colour scales of the images are from -0.03 to 0.03 in linear scale
  for (a) and (c) and from 0.2 to 1.0 in logarithmic scale for (b).
  Colour bars are shown at the bottom of the figure.
}
\label{fig:destripe_map}
\end{figure}

The derived small-scale map was then Fourier-transformed to obtain a
two-dimensional spatial frequency map.
We applied the stripe cleaning process to the images in Ecliptic
coordinates, in which the scanning direction of the {\em AKARI} survey
is parallel to the y-axis of the images (see $\S$\ref{sec:observation}
for the spatial scan during the all-sky survey observations).
This is beneficial because the direction of the spatial stripe patterns caused by the spatial scans become
parallel to the y-axis, then noises caused by the stripes are confined
along the x-axis of the two-dimensional spatial frequency map as shown
in Figure~\ref{fig:destripe_map}~(a).
The stripe cleaning can be achieved by suppressing the power excess along the
x-axis.
While \citet{IRIS} made substitution of the pixel values in their
noise-contaminated regions with the azimuthal-averaged values, we
multiplied the spatial frequency map by
an attenuation factor $A(k)$, where $k$ is a wavenumber vector, to suppress the power excess.
This attenuation factor is similar to a filter in signal processing,
which is used for converting an input spectrum $F(k)$ to an output spectrum $G(k) = A(k)F(k)$.
Since a discontinuous filter function such as box-car function
brings ripples on the result of Fourier transform,
a smooth function is better for suppressing artefact.
Therefore, we calculated the attenuation factor as follows.
First we obtained a power spectrum $P(r,\theta)$,
where $(r,\theta)$ is a position on the two-dimensional spatial frequency map in
the polar coordinate system.
Next, we calculated the azimuthal average of $P(r,\theta)$ as $\bar{P_a}(r)$,
where a region around the x-axis was excluded from the averaging.
Then, we calculated the local average of $P(r,\theta)$ as $\bar{P_l}(r,\theta)$,
in the range of $r \pm \sqrt{r}/2$ [pixels] along the radial direction
of the spatial frequency map.
Finally, the attenuation factor was calculated as:
\[
  A(r,\theta) = \left\{
  \begin{array}{ll}
    [\bar{P_a}(r) / \bar{P_l}(r,\theta)]^{\frac{1}{2}} & ({\rm if}\ \bar{P_l}(r,\theta) > \bar{P_a}(r) \\
    &\ \ {\rm and}\ (r,\theta) \in R_A) \\
    1 & ({\rm otherwise})
  \end{array} \right.
\]
where $R_A$ is a region around the x-axis defined by a particular range in $\theta$ and $y$.
Figure~\ref{fig:destripe_map}~(b) shows an attenuation factor map obtained for a two-dimensional spatial
frequency map shown in Figure~\ref{fig:destripe_map}~(a).
Since $\bar{P_l}(r,\theta)$ is the local average of $P(r,\theta)$,
$A(r,\theta)$ has smooth distribution in the radial direction.
Figure~\ref{fig:destripe_map}~(b) also shows that
$A(r,\theta)$ is close to unity if $|y| > 0.05$ [arcmin$^{-1}$],
which means that our destripe process is not sensitive to the definition of $R_A$.
The resultant destriped spatial frequency map is shown in
Figure~\ref{fig:destripe_map}~(c), which is obtained by multiplying the spatial frequency map (a)
by the attenuation factor (b).
The result shows that the spatial frequency is attenuated
down to the level of the azimuthal average.

\begin{figure}
 \begin{center}
   \FigureFile(80mm,120mm){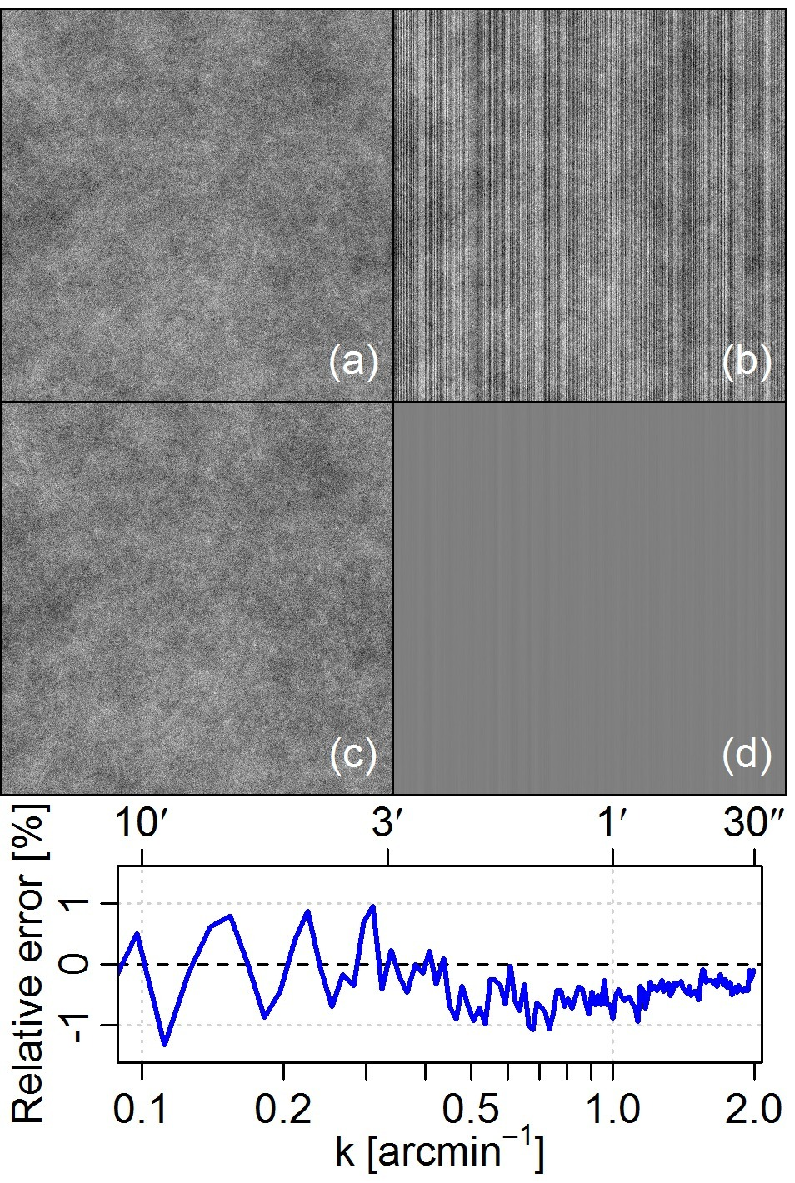}
 \end{center}
\caption{A verification of the image restoration with our destriping
  method. 
Panel (a) -- A synthetic image of ISM with random noise.
Panel (b) -- An input image for destriping made by adding stripe
noises to (a).
Panel (c) -- A destriped image.
Panel (d) -- Residual error of (c) minus (a).
The colour scales are same for the four images.
Bottom panel -- Relative error between the spatial spectra of (c) and
(a) derived by $[({\rm c}) - ({\rm a})] / ({\rm a})$.
}
\label{fig:destripe_simulation}
\end{figure}

We tested our destriping method in a similar way as \citet{IRIS}, using simulated images composed of the ISM with
a spectral power-law index of $\gamma = -3$ as well as random and
stripe noises.
The result is shown in Figure~\ref{fig:destripe_simulation}. Our simulation shows that
our method does not modify the power spectrum by more than 1\% at
small scales.

\subsection{Recursive deglitching and flat-fielding by referring processed image}\label{sec:deglitching}

Stripes were eliminated from the preliminary image by the
above mentioned process, which was a consequence of long-term sensitivity
drift of the detectors.
To restore those data and to perform better deglitching, we reprocessed the time-series data by referring against the preliminary image.

An additional gain and offset adjustment was made for each set of time-series data by each detector pixel to get better correction of the flat-fielding, by fitting each set to the preliminary image.
At the same time, outlier data rejection was also made by eliminating the data that exceed 10 times of the standard deviation from the reference image.

The final image was then processed from the time-series data and was cleaned by the image destriping process described above.

We processed the whole sky image in $6\degree \times 6\degree$ image patches with 5-degree
separations. These processed images were then combined into larger
images using the Montage software package (\cite{Berriman08}).

\section{Results}\label{sec:results}

\begin{figure*}
  \begin{center}
   \FigureFile(160mm,220mm){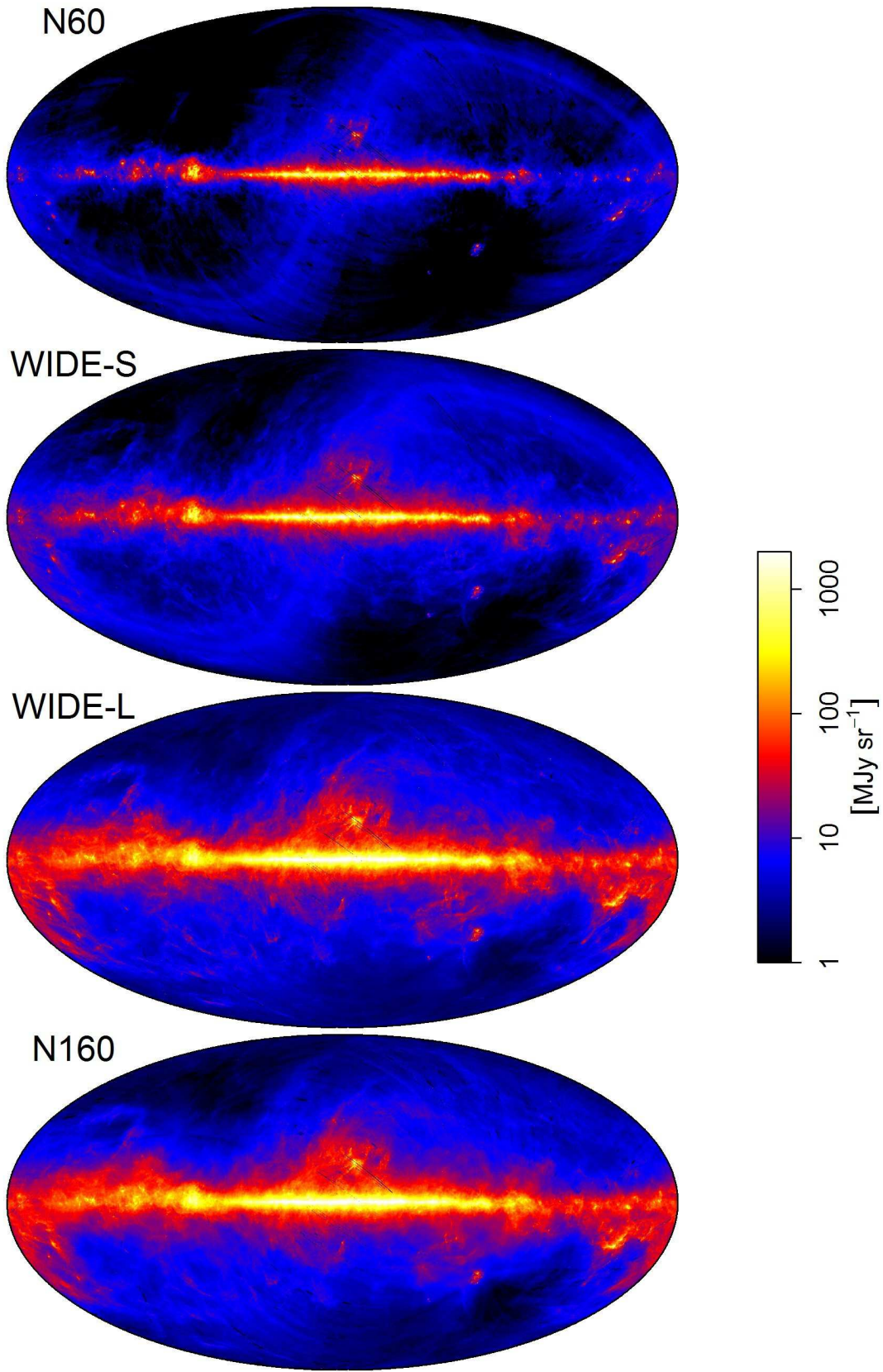}
  \end{center}
  \caption{All-sky FIR intensity maps observed by {\it AKARI} survey.}
\label{fig:4plates}
\end{figure*}

\begin{figure*}
  \begin{center}
   \FigureFile(160mm,122mm){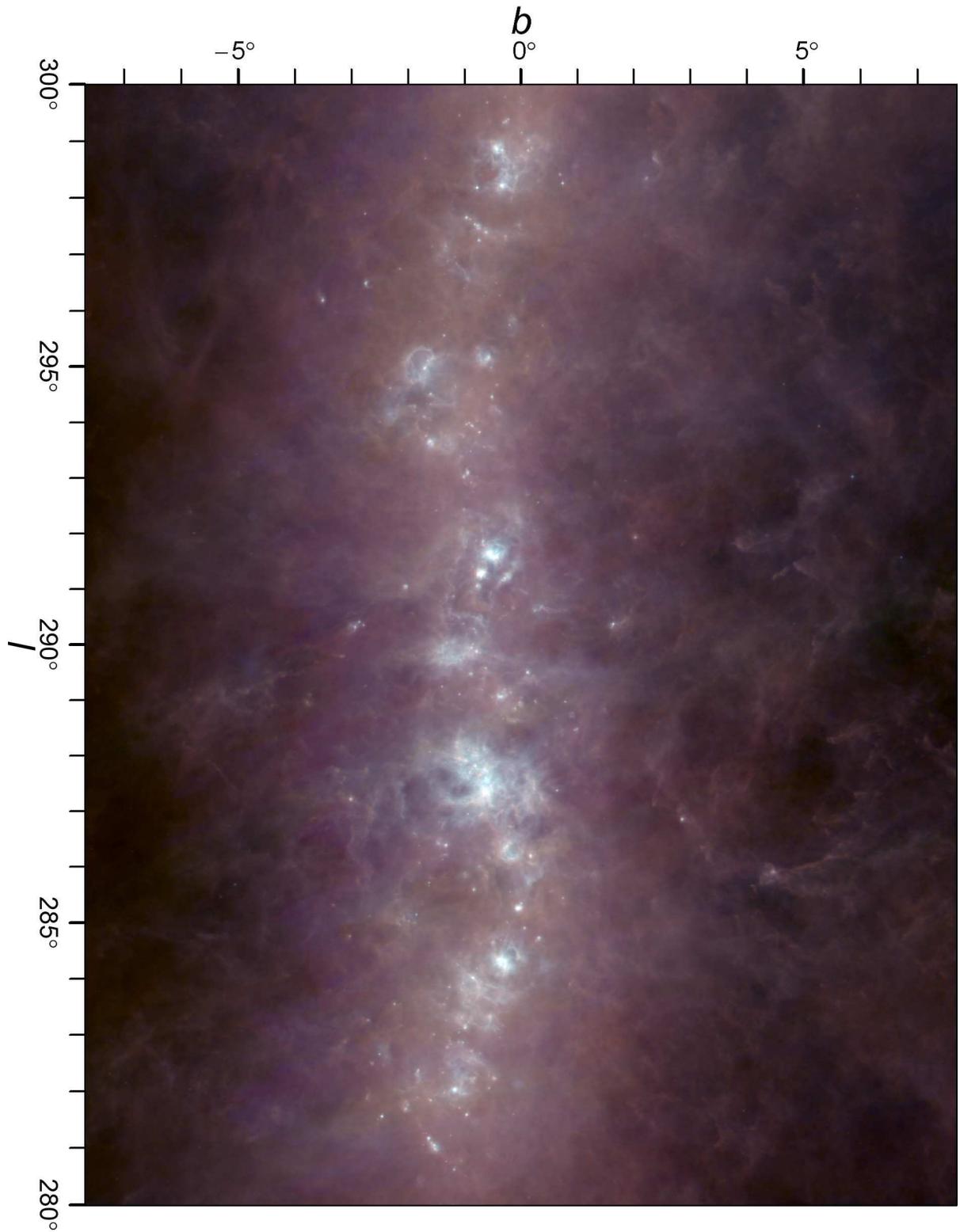}
  \end{center}
  \caption{A three-colour composite image of the Galactic plane around
    $\eta$ Carinae region. N60 (blue), WIDE-S (green), and WIDE-L (red) images
    are combined. Intensities are from 1 to 10000 [MJy sr$^{-1}$] in
    logarithmic scale.}
\label{fig:etaCar}
\end{figure*}

Intensity maps from the all-sky survey in the four photometric bands are shown in Figure~\ref{fig:4plates}.
Detailed photometry with high spatial resolution has been achieved
for the whole sky.
The far-infrared intensity distribution of the emission shows a clear wavelength
dependence, concentrating more to the Galactic plane at shorter wavelengths with significant extension towards higher galactic latitudes being seen at longer wavelengths.
The concentration to the Galactic plane indicates a tighter connection
to the star-formation activity by tracing high ISRF regions at and around
star-formation regions.
The extended emission seen at longer wavelengths traces spatial
distribution of low temperature dust in high latitude cirrus clouds.
This noticeable dependence of the spatial distribution on the
wavelength difference between 65 \micron\ and 160 \micron\ shows the importance of the wide wavelength coverage of {\it AKARI} to observe the FIR dust
emission at the peak of its SED where it has the largest dependence on
the temperature difference of the dust particles.
This difference in concentration to the Galactic plane is also visible
in a zoom-up image of the Galactic plane at $l = 280\degree$ --
300\degree\ (Figure~\ref{fig:etaCar}).

\begin{figure}[ht]
  \begin{center}
   \FigureFile(80mm,160mm){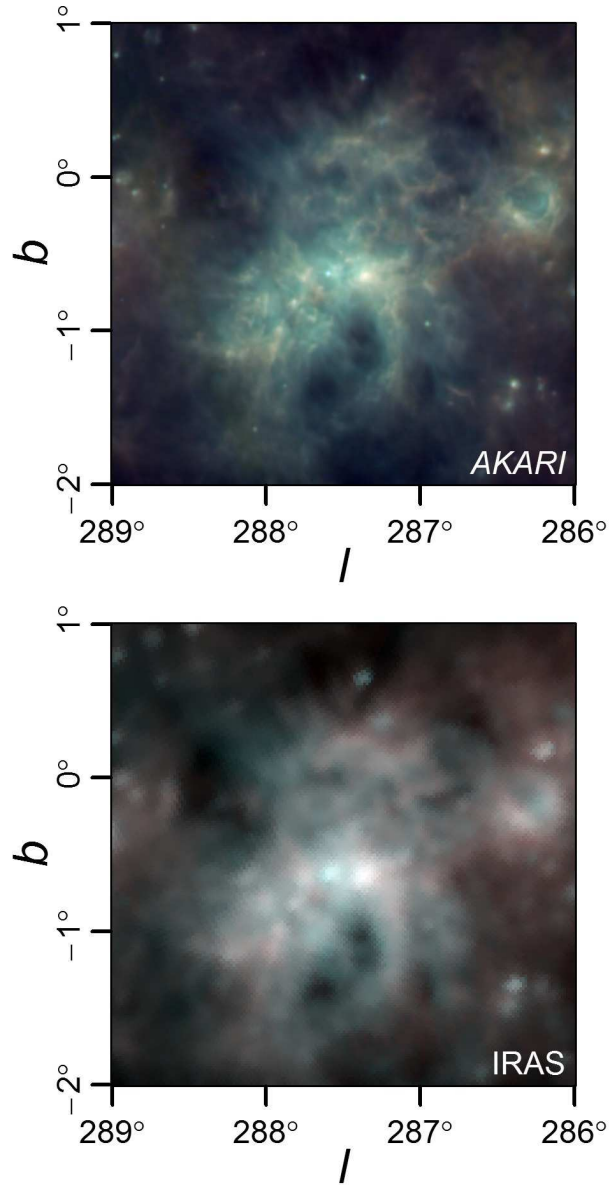}
  \end{center}
  \caption{Zoom-up images of the $\eta$ Carinae
    region. Upper panel -- a three-colour composite of N60 (blue), WIDE-S (green), and
    WIDE-L (red) images. Lower panel -- a composite image of IRIS 60
    \micron\ (cyan) and IRIS 100 \micron\ (red) images for
    a comparison of spatial resolutions between IRAS and {\it AKARI} images.}
\label{fig:etaCarzoom}
\end{figure}

The emission ranging from the tenuous dust in high Galactic cirrus
clouds to the bright inner galactic plane, has been successfully
measured in the {\it AKARI} all sky images.
A companion paper (\cite{Takita14}) describes further details of the data calibration and the reliability estimation, and confirms that a good
linearity correction has been achieved for all the bands, from the
faintest detector signals up to the bright signals of $>10$ [GJy
sr$^{-1}$], with relative accuracy of $\sim 10\%$.
It is important to note that the good linearity correction achieved
for the data also confirms the successful correction of the spatial
distortion of the data caused by the transient response of the
detector.
The intensity calibration is done primarily by using more sensitive
slow-scan data (\cite{Shirahata09}), and the linearity and absolute
intensity have been checked with IRAS/IRIS (\cite{IRIS}) and
COBE/DIRBE (\cite{Hauser98}).
\citet{Takita14} estimates the absolute accuracy of 20\% has been
achieved for all the bands with intensities of $\geq6$ [MJy sr$^{-1}$]
for N60, $\geq2$ [MJy sr$^{-1}$] for WIDE-S, and $\geq15$ [MJy sr$^{-1}$]
for WIDE-L and N160.
The better absolute accuracy has been achieved for brighter regions,
and that of 10\% has been achieved for $\geq10$ [MJy sr$^{-1}$]
for N60, $\geq3$ [MJy sr$^{-1}$] for WIDE-S, $\geq25$ [MJy sr$^{-1}$]
for WIDE-L, and $\geq26$ [MJy sr$^{-1}$] for N160.
\citet{Takita14} also estimates the full width at half maximum of
point spread function as 63\arcsec, 78\arcsec, and 88\arcsec\ for N60,
WIDE-S, and WIDE-L. Although no evaluation of the point spread function is
available for N160 due to the lower sensitivity of the N160 band
(\cite{Takita14}), comparable point spread function is expected for
N160 as WIDE-L referring to the more sensitive slow-scan data
(\cite{Shirahata09}).
Improvement of the spatial resolution comparing to IRAS images is
demonstrated in Figure~\ref{fig:etaCarzoom}.

The residual of the Zodiacal emission (ZE) along the
ecliptic plane is visible in shorter wavelength images, especially in
the N60 image.
This is because we have subtracted the spatially smooth component from
the data but have not subtracted the dust band component, which has
smaller scale structures that are for the first time revealed by the
high spatial resolution {\it AKARI} observation ($\S$\ref{sec:zodi}).
The former models of the ZE cannot reproduce this
spatial distribution properly so that we need to develop our own model
to remove the emission from the celestial image.
Our current estimation of the ZE is briefly described
in $\S$\ref{sec:cavzodi} and will be investigated in detail in a future paper \citep{Ootsubo14}.

\section{Discussion}\label{sec:discussion}

\subsection{Spatial Power Spectra}

Our survey images are the first-ever FIR images that cover the whole sky with arc-minute spatial resolution.
A major advantage of the {\it AKARI} data is that it enables us to obtain the global distribution of the ISM with higher spatial resolutions.
In other words, the large spatial dynamic range of the data is one of
the key characteristics of the {\it AKARI} FIR survey.
This characteristic can be examined by calculating the spatial power spectra of the cirrus distribution taken from the {\it AKARI} all sky survey, as the cirrus power spectra can be well represented by power-law spectra ($\S$\ref{sec:intro}).

\citet{mamd10} derived the IRAS/IRIS 100 \micron\ and Herschel/SPIRE 250 \micron\ combined spatial power spectrum at the Polaris flare region and confirmed that the power-law nature of the cirrus power spectrum is kept down to sub-arcminute scales, with a power-law index $\gamma = -2.65 \pm 0.10$ on scales $0.025 < k < 2 {\rm\ [arcmin^{-1}]}$ (also see \cite{Martin10}).
Since the Polaris flare is a high Galactic latitude cirrus cloud with
virtually no sign of star-formation activity
(\cite{2010A&A...518L..92W}; \cite{Martin10}), the region is well suited to assess the nature of the cirrus power spectrum.

\begin{figure*}
  \begin{center}
    \FigureFile(160mm,230mm){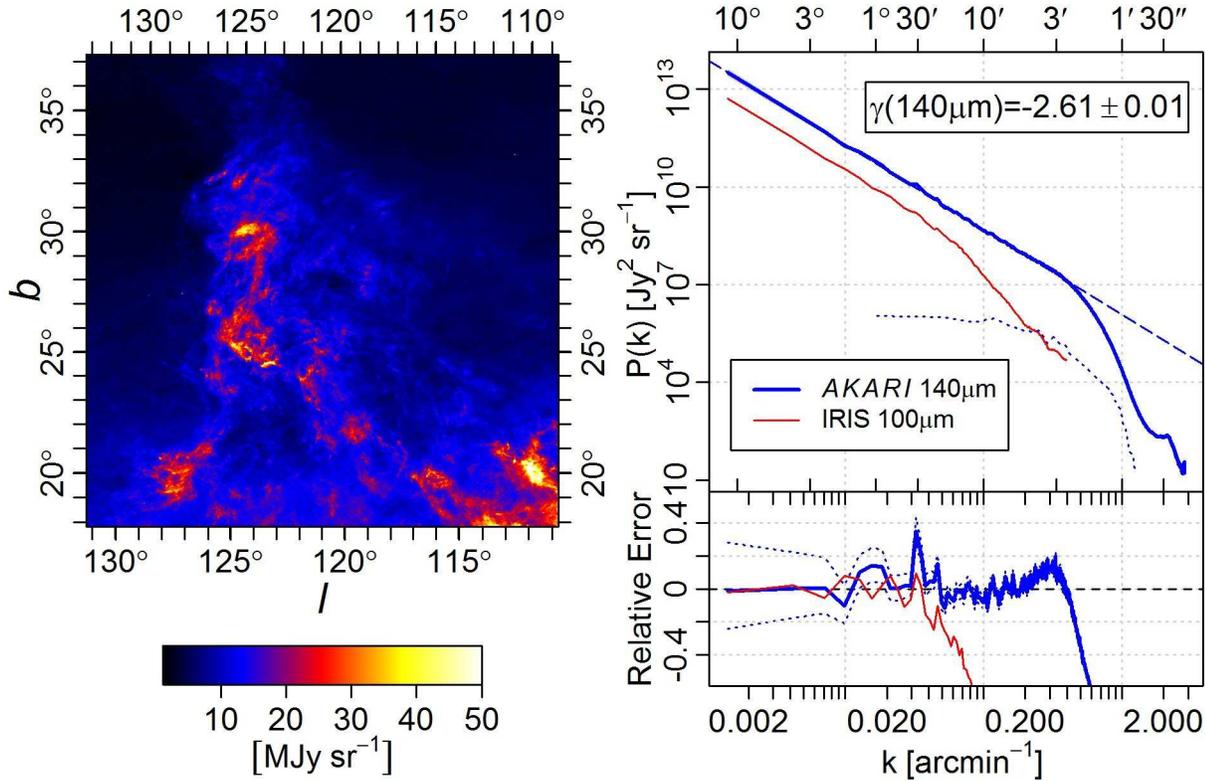}
  \end{center}
  \caption{Left panel -- WIDE-L intensity image of the Polaris Flare region in
    Galactic coordinates. Right upper panel -- Spatial power spectrum of WIDE-L
    intensity (the blue solid line) observed in the Polaris flare region shown in the left
    figure. A power-law fitting of the spectrum at the scales of
    $0.0014 < k < 0.21 {\rm\ [arcmin^{-1}]}$ is shown as the blue dashed
    line with the power index of $-2.61 \pm 0.01$.
    The blue dotted line shows the spectrum of the point spread
    function, which is arbitrary shifted in the vertical direction for the
    comparison with the WIDE-L power spectrum.
    The red solid line shows the spatial spectrum of the IRIS
    100 \micron\ data of the same region for comparison.
    Right lower panel -- Relative error of WIDE-L spatial spectrum from the power-law fitting shown
    in the right upper panel (the blue solid line). Standard error
    of the WIDE-L spatial spectrum estimation is denoted as the blue
    dotted line. Relative error of
    the IRIS 100 \micron\
    data is also shown as the red solid line for comparison.
}
\label{fig:PolarisFlare}
\end{figure*}

The {\it AKARI} FIR image of the Polaris flare region and its spatial
power spectra are shown in Figure~\ref{fig:PolarisFlare} (also see \cite{2012PKAS...27..111D}).
Good linearity is found from a small scale of $\sim3'$ to the larger
scales beyond 10\degree, which is consistent with the measurements of
many other authors (\cite{mamd10} and the references therein).  The
power law index is estimated as $\gamma_{140 \micron} = -2.61 \pm
0.01$ by fitting the WIDE-L 140 \micron\ spectrum in $k = 0.0014 $ --
$ 0.21$ [arcmin$^{-1}$] wavenumber range.  An excess component from
the fitted power-law spectrum is found in the wavenumber range $\sim
0.2 - 0.4$ [arcmin$^{-1}$] (the right lower panel of the
Figure~\ref{fig:PolarisFlare}; also see Figure~3 of
\cite{2012PKAS...27..111D}), which is due to the noise in the data.
So we limit the wavenumber range for the fit as $<0.21$ [arcmin$^{-1}$] and eliminate
all the data above this wavenumber limit.  This excess component
should be negligible in the power spectra of brighter regions.  The
deviation from the power-law spectrum at the scales smaller than $\sim
3\arcmin$ can be attributed to the spatial resolution of the WIDE-L
band, whose point spread function (PSF) is shown as the dotted line in
the Figure~\ref{fig:PolarisFlare}.

The estimated $\gamma$ is in good agreement with that by
\citet{mamd10} shown above.
The spectra clearly show the wide dynamic range of our observation, as
the spatial power component is well retrieved from $> 10^{\circ}$
large scale distribution to $< 5'$ small scale structures.

One advantage of the {\it AKARI} FIR data compared to the IRAS data is illustrated in Figure~\ref{fig:PolarisFlare}, which shows deviation from the power-law distribution well above the spatial scales that the {\it AKARI} FIR data shows the deviation to be, due to the improved spatial resolution of the {\it AKARI} data.
The advantage of the improved spatial resolution of the {\it AKARI}
FIR data is thus clearly indicated.

\subsection{Filamentary Structure}

\begin{figure*}
  \begin{center}
    \FigureFile(160mm,140.0mm){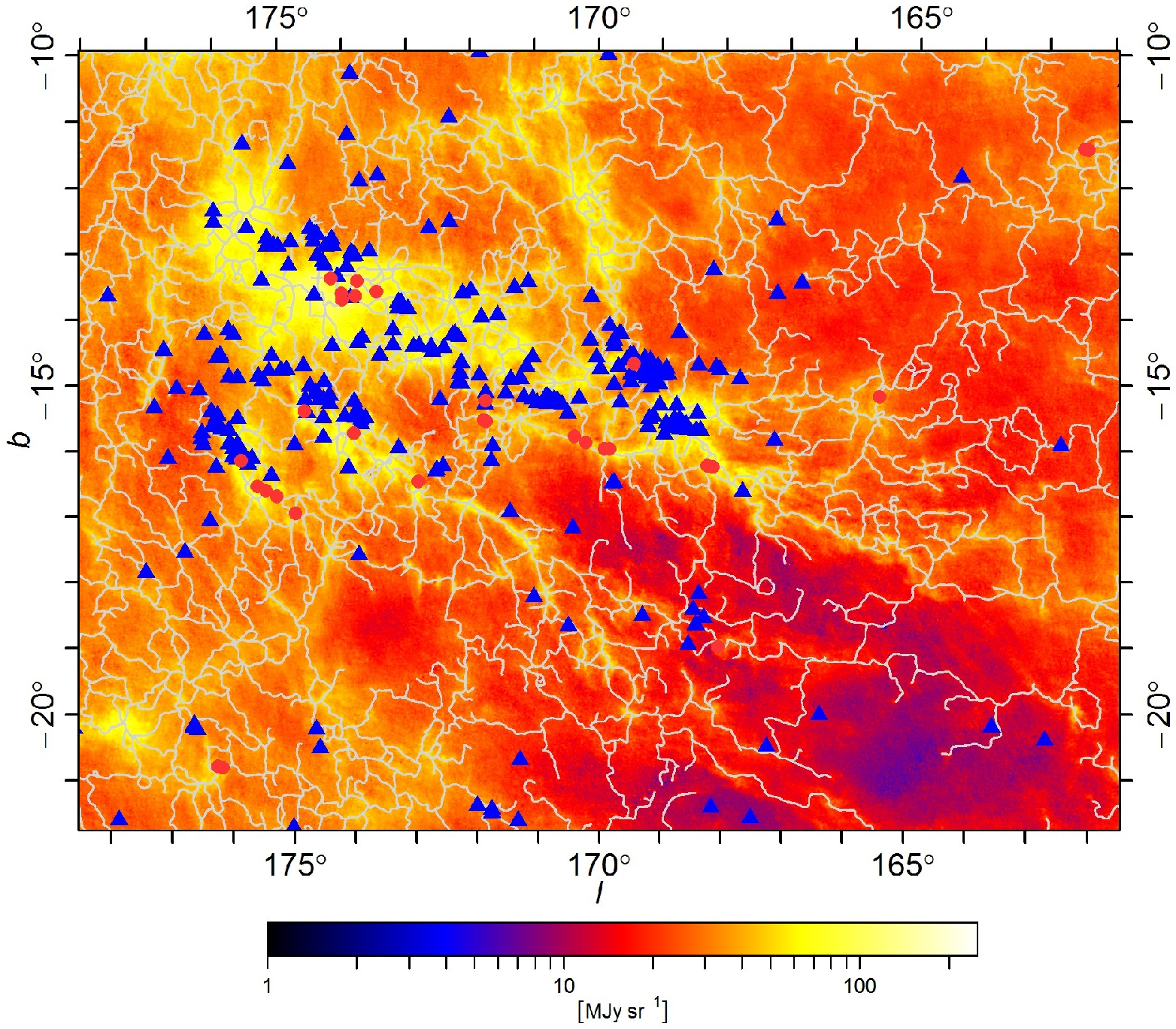}
  \end{center}
  \caption{Filamentary structures extracted from the {\it AKARI} all-sky survey image of the Taurus region (\cite{Doi14}). The skeleton of the filamentary structures, which is extracted from the {\it AKARI} image by applying DisPerSE algorithm (\cite{2013ascl.soft02015S}), is superposed on the {\it AKARI} WIDE-L all-sky survey image.
The red circles are 30 candidate sources of young stellar objects identified in the {\it AKARI} bright source catalogue (\cite{2014PASJ...66...17T}).
The blue triangles are 303 known T-Tauri stars in the region
(\cite{2010A&A...519A..83T}), a compilation of published catalogues by
\citet{1989AJ.....97.1451S}, \citet{1990AJ.....99..924B},
\citet{1996A&A...312..439W}, \citet{1997A&AS..124..449M},
\citet{1998A&AS..132..173L}, \citet{2007A&A...468..353G},
\citet{2008hsf1.book..405K}, and \citet{2010ApJS..186..259R}.
 }
  \label{fig:TauFilament}
\end{figure*}
\begin{figure}[ht]
  \begin{center}
    \FigureFile(80mm,80mm){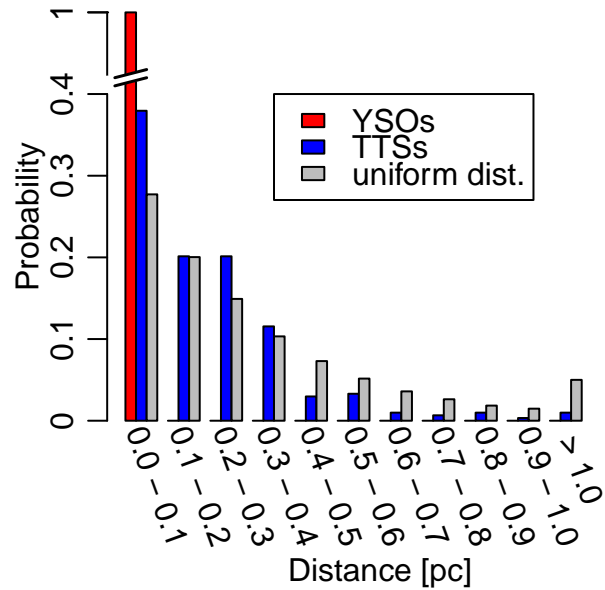}
  \end{center}
  \caption{A histogram of the distance from young stellar objects (YSOs; red bars) and T-Tauri stars (TTSs; blue bars) in the region shown in Figure~\ref{fig:TauFilament} to their nearest filamentary structures.
Distance to the Taurus region is assumed as 137 pc (\cite{2007ApJ...671.1813T}).
All the 30 YSOs show spatial coincidence with the extracted filaments within the range of our spatial resolution ($<0.05$ pc).
TTSs show moderate concentration to the filaments with the distance $<0.4$ pc.
The gray bars show an expected distribution of the uniformly distributed sources in the region.
The significance of the difference between the distribution of TTSs and that of uniformly distributed sources is estimated as p-value $ = 0.004$ by the chi-square test.
}
  \label{fig:FilamentDistance}
\end{figure}
\begin{figure}[ht]
  \begin{center}
    \FigureFile(80mm,80mm){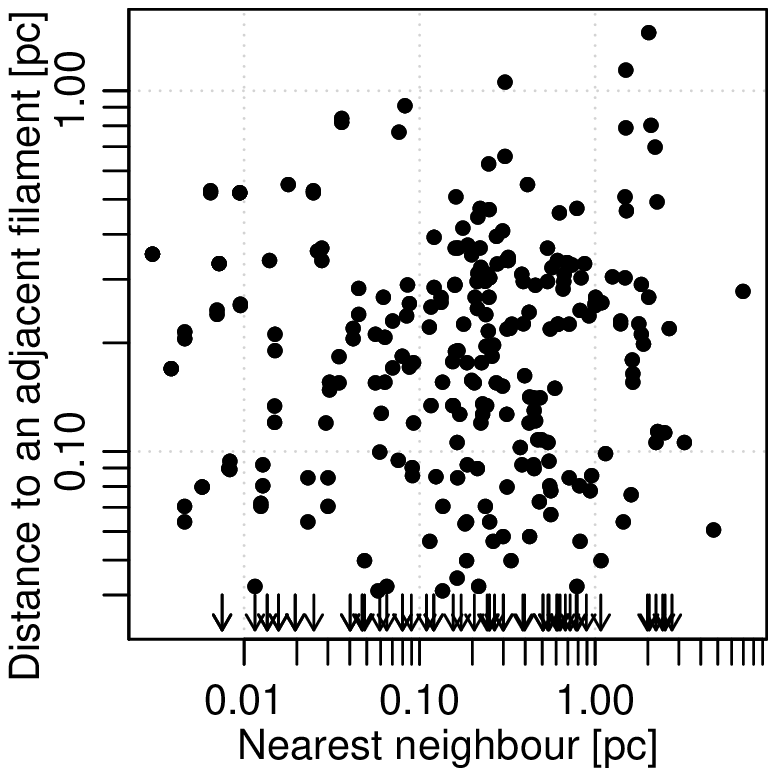}
  \end{center}
  \caption{Correlation between the distance of the TTSs to their nearest neighbour TTSs and the distance of the TTSs to their adjacent filaments. Arrows are the sources who show positional correspondence to filaments within our spatial resolution. The nearest neighbour distance is an indicator of the crowdedness of the TTSs' distribution. We find no dependence of the TTSs' distance to their adjacent filaments whether they are in the crowded regions or in an open field.}
  \label{fig:TTScrowdedness}
\end{figure}
Recent observations by the Herschel satellite shows that cirrus clouds consists from filamentary structures, whose typical width is $\sim 0.1$ pc ($\S$\ref{sec:intro}).
This typical width corresponds to $3'\negthinspace.4$ -- $1'\negthinspace.1$ at the distance of 100 -- 300 pc from the sun.
Thus these filaments in the local clouds are detectable in the {\it AKARI} images, and the global distribution of the filamentary structures can be revealed by the {\it AKARI} all-sky survey (\cite{Doi14}).
Figure~\ref{fig:TauFilament} shows an example of the filamentary structure extraction by applying DisPerSE algorithm (\cite{2013ascl.soft02015S}) to an {\it AKARI} image of the Taurus molecular cloud (\cite{Doi14}).
Ubiquitous distribution of the filamentary structure is displayed.

We also plot candidate sources of young stellar objects (YSOs; \cite{2014PASJ...66...17T}) and known T-Tauri stars (TTSs; \cite{2010A&A...519A..83T}) in Figure~\ref{fig:TauFilament} to check their spatial correlation with the filamentary structures (also see Figure~\ref{fig:FilamentDistance}).
All the YSO candidates (30 out of 30 sources) in the region show spatial coincidence with the filamentary structure within the range of our spatial resolution.
This correspondence is consistent with the Herschel observations ($\S$\ref{sec:intro}), who found $> 70\%$ of prestellar cores in the filaments (\cite{2014prpl.conf...27A}), indicating a strong connection of the filamentary structures and star-formation activities.
TTSs show less concentration to the filaments, but still significantly
higher concentration than that expected for uniformly distributed
sources in the region with a p-value of 0.004.
Figure~\ref{fig:TTScrowdedness} shows the correlation between the
distance of the TTSs to their nearest neighbour TTSs, which we take as
an indicator of the crowdedness of the TTSs' distribution, and the
distance of the TTSs to their adjacent filaments.
Since we find no correlation in the plot, we conclude that the moderate
concentration of the TTSs found in
Figures~\ref{fig:TauFilament} \& \ref{fig:FilamentDistance} is not because that TTSs tend to distribute in the dense cirrus regions, which have crowded filamentary structures.
This moderate concentration of the TTSs can be explained as an age evolution of their distribution. A proper motion of 0.1 [km s$^{-1}$] of the TTSs against their natal filaments for 1 -- $5\times 10^6$ years gives distance of 0.1 -- 0.5 [pc], which is consistent with our observed projected distance of the TTSs from their adjacent filaments.

The large spatial dynamic range of the {\it AKARI} observation enables us to reveal global distribution of the filamentary structures, and to study their correlation with young stellar sources of various evolutionary stages.

\subsection{Spectral Energy Distribution}

\begin{figure*}
  \begin{center}
    \FigureFile(160mm,101.6mm){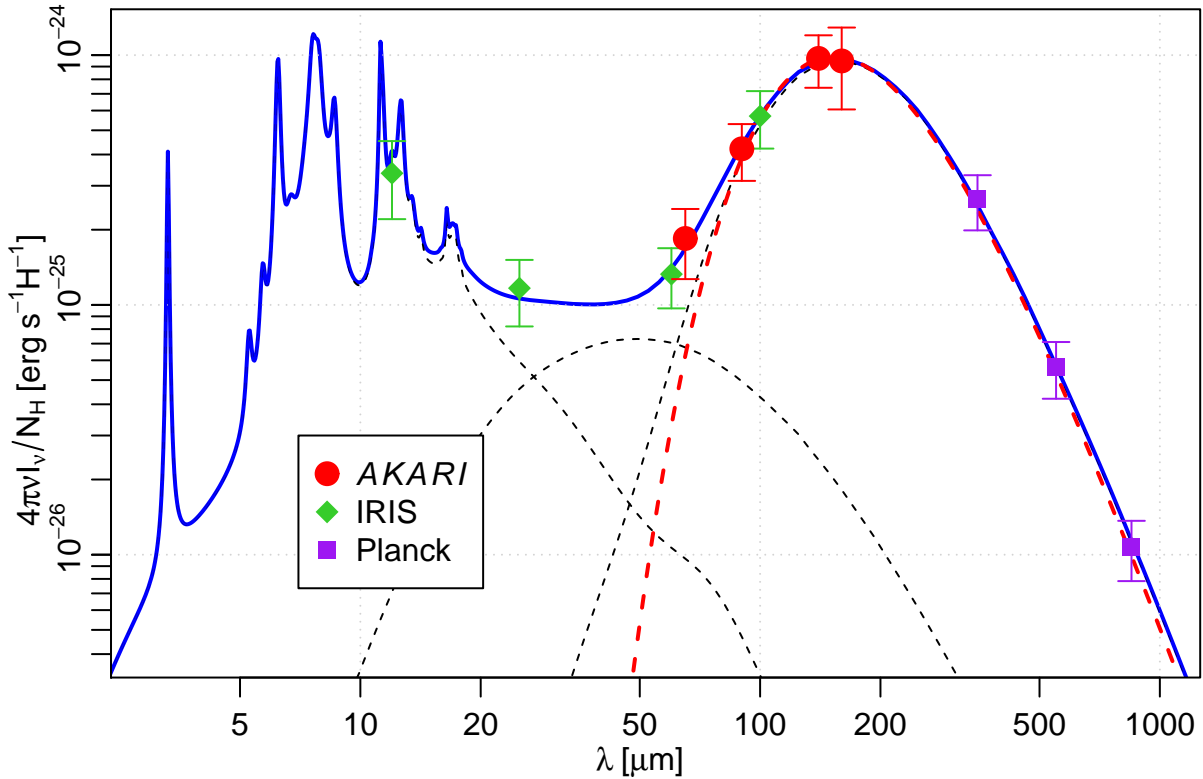}
  \end{center}
  \caption{Spectrum Energy Distribution at the Polaris Flare region
    shown in Figure~\ref{fig:PolarisFlare} by fitting the spatial
    power spectra of each observational band by assuming the power low
    index of -2.61, which is estimated for the WIDE-L data.
The square root of the relative amplitude of the spatial power spectra and
their fitting errors are shown for the four {\it AKARI} FIR data as well as the ancillary IRIS and Planck
    data.
    The blue solid line indicates a model fitting of the four {\it AKARI}
    FIR data by applying DustEM SED model (\cite{Compiegne11}).
    The black dotted lines are the decomposition of the modelled SED in
    BGs, SGs, and PAHs.
    The relative fraction of the excess emission from SGs against that
    from BGs estimated from the fitted model SED are 70.0\% for N60, 16.4\% for
    WIDE-S, 2.6\% for WIDE-L, and 2.0\% for N160.
    The red dashed line indicates the estimation of colour temperature
    ($T = 16.8$ [K]) from the intensity ratio between WIDE-S and WIDE-L data
    assuming a modified black-body spectral index $\beta = 1.62$.
  }
  \label{fig:sed}
\end{figure*}

The SED of the cirrus emission can be estimated from the square root of
the amplitude of the spatial power spectra (\cite{Roy10}).
We estimate the wavelength dependence of the amplitude in the Polaris
Flare region, whose area is shown in Figure~\ref{fig:PolarisFlare}.
The power law index of the spectrum scales is assumed as $\gamma =
-2.61$ for all the four {\it AKARI} FIR bands as well as the ancillary
data of IRIS and Planck.
The estimated relative amplitudes and their fitting
errors are shown in Figure~\ref{fig:sed}.

In addition to the spatial dynamic range of the {\it AKARI} data, another key aspect is its ability to sample across the spectral peak of the dust SED with four photometric wave-bands.
To demonstrate this aspect, we performed a model fit to the {\it
  AKARI} FIR data
with the \citet{Compiegne11} dust model, using the DustEM numerical tool (http://www.ias.u-psud.fr/DUSTEM/), as indicated in Figure~\ref{fig:sed}.
It is clear that we can accurately reproduce the FIR dust SED with the {\it AKARI} FIR data, as the fitted model SED agrees well
with all the ancillary data.

The colour temperature from the WIDE-S / WIDE-L intensity ratio (90
\micron\ / 140 \micron) is estimated to be $T = 16.8$ [K].
A modified black body spectrum of the estimated temperature is shown
in Figure~\ref{fig:sed} as the red dashed line.
A modified black-body spectral index is assumed as $\beta = 1.62$,
which is a mean value over the whole sky estimated by
\citet{2014A&A...571A..11P} by fitting IRIS 100 \micron\ and
Planck 353, 545, and 857 GHz data with modified black body spectra.
The estimated colour temperature spectrum is consistent with the FIR
-- sub-mm part of the BGs' emission spectrum as well as the Planck
observations.

The {\it AKARI} FIR data is therefore a good tracer of the temperature, and the
total amount of BGs, and as a result of the ISM as a whole.
Together with the much improved spatial resolution from the former
all-sky survey data in the FIR wavebands, the newly achieved {\it AKARI} FIR high-spatial resolution images of the whole sky should be a powerful tool to investigate the detailed spatial structure of ISM and its physical environment.

\section{Caveats and plans for future improvements}\label{sec:caveats}

Although we have produced a sensitive all sky FIR images, the data
still contains some artefacts, and there is scope to make further improvements.
In the following, we describe remaining caveats about the efficacy of
the data and our plans to mitigate these remaining problems.

\subsection{Zodiacal emission}\label{sec:cavzodi}

Although we have subtracted the smooth cloud component of the ZE
from the raw data, we have not yet subtracted the asteroidal
dust band component, nor the MMR component from the data, with the
consequence that the contributions from these remain in the images, and
is recognisable at the shorter wavelength bands (N60 and WIDE-S).
This is because small scale structures of the ZE are not adequately
reproduced by the former Zodiacal emission models ($\S$\ref{sec:zodi}).
Asteroidal dust bands appear as pairs of parallel bands
equally spaced above and below the ecliptic plane.
The continuous and smooth distribution of the dust bands and
the MMR component along the ecliptic plane
can be seen in the {\it AKARI} N60 and WIDE-S maps (Figure~\ref{fig:4plates}).

\begin{figure}[ht]
  \begin{center}
    \FigureFile(80mm,80mm){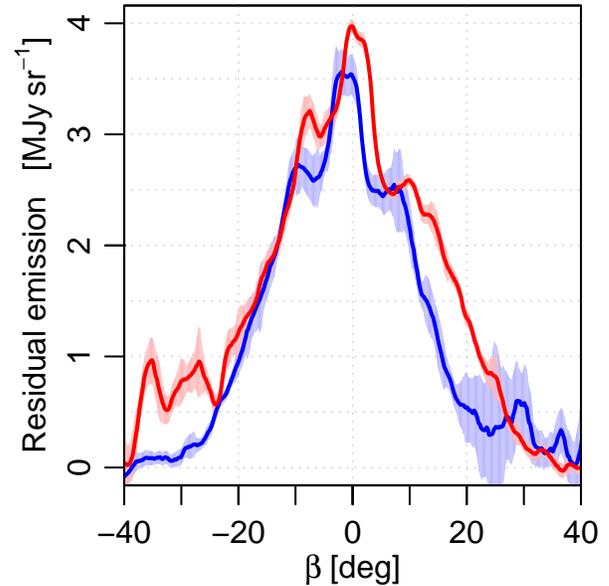}
  \end{center}
  \caption{
Example latitudinal profiles of the dust bands and the MMR component
observed in the WIDE-S band in two regions,
$175\degree < \lambda < 185\degree$ (the solid red line) and
$-5\degree < \lambda < 5\degree$ (the solid blue line).
The shaded areas denote the standard error of the profile.
The intensity of the residual ZE contribution in the AKARI map are
less than 5 [MJy sr$^{-1}$] at the shorter wavelength bands.
}
  \label{fig:zodiprofile}
\end{figure}

The intensity and the ecliptic latitudes of the peak positions
of these components change depending on the ecliptic longitude.
We briefly evaluate the spatial distribution and the latitudinal profiles
of these components at the shorter wavelength bands, N60 and WIDE-S.
The contribution of the asteroidal dust bands and the MMR component will be
discussed in more detail in \citet{Ootsubo14}.
Figure~\ref{fig:zodiprofile} shows example latitudinal profiles of the residual ZE contribution from
the dust bands and the MMR component observed in the WIDE-S band in two
regions,
where the ZE contributions are strong ($175\degree < \lambda < 185\degree$ and $-5\degree < \lambda < 5\degree$).
The estimated intensities of the emission that are left in the {\it
AKARI} FIR images become maximum near the ecliptic plane and are $< 5$ [MJy sr$^{-1}$] for N60 and $< 4$ [MJy sr$^{-1}$] for WIDE-S, respectively.
Although the ZE contribution cannot be clearly seen in the WIDE-L and N160 images,
the estimated intensities are $< 1$ [MJy sr$^{-1}$] for WIDE-L
and N160 at most, if we assume that these components
have the dust temperature $T = 200$ -- 300 K.

\subsection{Moving bodies}\label{sec:cavmoving}

Planets and asteroids have not been masked during our image processing as the
detection of the faint sources are not fully investigated and the
scattering pattern of the bright sources are yet to be analysed.
So the images near the ecliptic plane may contain these solar system
bodies and will need further consideration in the future.
Meanwhile, we list positions of the planets and 55 major asteroids
that are scanned during the {\it AKARI} FIR all-sky survey observation
in Table~\ref{tab:moving}.

\begin{longtable}{lrrrrrr}
  \caption{List of planets' and 55 major asteroids' positions that are scanned during the {\it AKARI} FIR all-sky survey observation.}
  \label{tab:moving}
      \hline
      \vspace*{-3mm} &&&&&&\\
      & \multicolumn{2}{c}{Equatorial Coordinates} & \multicolumn{2}{c}{Galactic Coordinates} & \multicolumn{2}{c}{Ecliptic Coordinates} \\
      Source name & \multicolumn{1}{c}{R.A. (J2000)} &
      \multicolumn{1}{c}{Dec. (J2000)} & \multicolumn{1}{c}{\it l} & \multicolumn{1}{c}{\it b} & \multicolumn{1}{c}{$\lambda$} & \multicolumn{1}{c}{$\beta$} \\
      & $^{\rm h}$~~~$^{\rm m}$~~~~$^{\rm s}$ & \degree~~~\arcmin~~~\arcsec & \degree~~~\arcmin~~~\arcsec & \degree~~~\arcmin~~~\arcsec & \degree~~~\arcmin~~~\arcsec & \degree~~~\arcmin~~~\arcsec \\
      \hline
      \vspace*{-3mm} &&&&&&\\
\endhead
      \vspace*{-3mm} &&&&&&\\
      \hline
\endfoot
      \vspace*{-3mm} &&&&&&\\
      \hline
\endlastfoot
Philomela & 0~~23~~20~ & 13~10~12~ & 113~27~40~ & -48~57~18~ & 10~36~32~ & 9~46~07~ \\
Interamnia & 0~~45~~40~ & 9~05~33~ & 121~37~09~ & -53~29~55~ & 14~02~59~ & 3~51~15~ \\
Emma & 1~~46~~58~ & 11~12~19~ & 144~56~03~ & -48~54~60~ & 28~50~02~ & 0~09~15~ \\
Juno & 1~~52~~50~ & 5~19~21~ & 150~37~56~ & -53~49~35~ & 28~06~34~ & -5~51~41~ \\
Aurora & 2~~02~~41~ & -5~16~59~ & 164~42~45~ & -61~46~38~ & 26~37~40~ & -16~39~05~ \\
Isis & 2~~11~~22~ & 8~30~20~ & 154~58~10~ & -48~55~39~ & 33~32~49~ & -4~27~20~ \\
Wratislavia & 2~~15~~31~ & 13~19~50~ & 153~00~36~ & -44~11~01~ & 36~06~47~ & -0~14~36~ \\
Fides & 2~~18~~36~ & 20~27~01~ & 149~44~12~ & -37~27~58~ & 39~09~44~ & 6~14~08~ \\
Bellona & 2~~36~~59~ & 9~14~59~ & 162~17~52~ & -44~58~36~ & 39~47~58~ & -5~47~36~ \\
Thisbe & 2~~49~~41~ & 13~44~55~ & 162~15~11~ & -39~30~26~ & 44~08~06~ & -2~26~35~ \\
Juno & 3~~03~~34~ & 28~51~30~ & 155~25~07~ & -25~09~50~ & 51~35~47~ & 11~06~23~ \\
Pallas & 3~~05~~26~ & 10~11~17~ & 169~02~57~ & -39~56~55~ & 46~49~11~ & -6~57~14~ \\
Athamantis & 3~~10~~38~ & 11~46~18~ & 168~58~07~ & -37~55~01~ & 48~29~17~ & -5~46~41~ \\
Uranus & 3~~14~~48~ & 23~18~07~ & 161~17~27~ & -28~16~60~ & 52~30~58~ & 5~04~25~ \\
Uranus & 3~~18~~39~ & 9~04~28~ & 173~08~19~ & -38~35~26~ & 49~41~16~ & -8~54~08~ \\
Herculina & 3~~23~~38~ & 22~21~32~ & 163~46~00~ & -27~46~52~ & 54~14~39~ & 3~38~18~ \\
Palma & 3~~35~~10~ & 32~19~34~ & 159~11~41~ & -18~23~14~ & 59~13~37~ & 12~40~18~ \\
Melpomene & 3~~36~~35~ & 5~07~32~ & 180~41~56~ & -38~06~12~ & 53~04~22~ & -13~50~19~ \\
Emma & 4~~06~~26~ & 19~17~60~ & 174~10~29~ & -23~10~55~ & 63~19~29~ & -1~33~04~ \\
Philomela & 5~~17~~30~ & 6~36~23~ & 195~59~57~ & -16~50~58~ & 78~59~35~ & -16~25~49~ \\
Euphrosyne & 6~~00~~57~ & 7~30~24~ & 200~43~43~ & -7~04~04~ & 90~14~49~ & -15~56~06~ \\
Hesperia & 6~~32~~01~ & 28~10~53~ & 185~55~37~ & 9~10~22~ & 97~04~39~ & 4~56~09~ \\
Thetis & 7~~14~~39~ & 22~26~34~ & 195~23~44~ & 15~30~15~ & 107~12~06~ & 0~06~30~ \\
Themis & 7~~50~~56~ & 10~22~05~ & 210~38~56~ & 18~25~14~ & 117~44~32~ & -10~26~33~ \\
Minerva & 7~~51~~35~ & 24~25~59~ & 196~49~42~ & 24~06~00~ & 115~15~31~ & 3~24~16~ \\
Lutetia & 8~~05~~09~ & 22~44~33~ & 199~46~51~ & 26~25~16~ & 118~38~33~ & 2~21~29~ \\
Diotima & 8~~11~~14~ & 18~49~09~ & 204~26~18~ & 26~19~50~ & 120~51~43~ & -1~10~35~ \\
Io & 8~~13~~18~ & 16~02~16~ & 207~30~29~ & 25~43~15~ & 121~57~01~ & -3~47~02~ \\
Callisto & 8~~18~~17~ & 20~57~56~ & 202~51~45~ & 28~39~19~ & 122~00~21~ & 1~16~59~ \\
Lutetia & 8~~21~~13~ & 9~51~47~ & 214~36~36~ & 24~55~15~ & 125~14~41~ & -9~21~50~ \\
Astraea & 8~~28~~39~ & 19~49~45~ & 205~04~20~ & 30~31~25~ & 124~37~52~ & 0~44~37~ \\
Nemausa & 8~~30~~12~ & 18~11~16~ & 207~00~02~ & 30~16~21~ & 125~23~06~ & -0~45~37~ \\
Thetis & 8~~46~~27~ & 39~54~50~ & 182~19~16~ & 38~59~22~ & 123~05~57~ & 21~08~01~ \\
Emma & 8~~48~~57~ & 45~19~03~ & 175~22~18~ & 39~44~09~ & 121~51~50~ & 26~26~16~ \\
Herculina & 8~~55~~19~ & 21~25~53~ & 205~48~29~ & 36~56~08~ & 130~14~57~ & 3~54~13~ \\
Loreley & 8~~55~~39~ & 15~40~07~ & 212~29~54~ & 34~57~38~ & 131~55~05~ & -1~36~48~ \\
Pallas & 9~~01~~57~ & 16~18~36~ & 212~28~57~ & 36~36~07~ & 133~11~22~ & -0~34~29~ \\
Jupiter & 9~~01~~60~ & 43~41~13~ & 177~32~14~ & 42~03~25~ & 124~52~60~ & 25~35~57~ \\
Minerva & 9~~03~~39~ & 12~07~55~ & 217~24~52~ & 35~18~02~ & 134~46~50~ & -4~27~39~ \\
Pluto & 9~~14~~34~ & 31~28~08~ & 194~26~06~ & 43~35~25~ & 131~27~15~ & 14~45~29~ \\
Metis & 9~~17~~31~ & 19~21~60~ & 210~32~14~ & 41~09~39~ & 135~50~15~ & 3~26~01~ \\
Amphitrite & 9~~24~~59~ & 16~32~36~ & 214~54~22~ & 41~48~08~ & 138~23~38~ & 1~17~23~ \\
Diotima & 9~~29~~38~ & 4~47~15~ & 229~13~33~ & 37~30~50~ & 143~11~19~ & -9~30~46~ \\
Hygiea & 9~~31~~55~ & 18~38~34~ & 213~03~26~ & 44~06~27~ & 139~18~21~ & 3~48~11~ \\
Doris & 9~~32~~60~ & 9~28~22~ & 224~30~58~ & 40~33~33~ & 142~28~33~ & -4~48~25~ \\
Metis & 9~~34~~18~ & 23~32~55~ & 206~41~22~ & 46~09~38~ & 138~15~01~ & 8~37~39~ \\
Victoria & 9~~37~~38~ & 13~37~39~ & 220~11~55~ & 43~26~03~ & 142~12~52~ & -0~30~27~ \\
Doris & 9~~43~~41~ & 17~06~07~ & 216~31~47~ & 46~09~36~ & 142~27~34~ & 3~15~11~ \\
Europa & 9~~45~~03~ & 12~33~42~ & 222~37~22~ & 44~36~15~ & 144~16~20~ & -0~55~23~ \\
Harmonia & 9~~49~~12~ & 14~17~53~ & 221~00~34~ & 46~15~40~ & 144~38~35~ & 1~02~58~ \\
Germania & 9~~55~~59~ & 3~33~08~ & 235~13~51~ & 42~22~05~ & 149~52~47~ & -8~29~32~ \\
Wratislavia & 10~~10~~52~ & 19~11~33~ & 217~07~28~ & 52~56~14~ & 147~49~51~ & 7~26~05~ \\
Patientia & 10~~27~~59~ & 14~33~07~ & 227~07~58~ & 54~49~17~ & 153~21~24~ & 4~35~27~ \\
Ino & 10~~34~~06~ & 4~57~30~ & 241~50~18~ & 50~50~31~ & 158~17~42~ & -3~46~23~ \\
Germania & 10~~37~~02~ & 3~02~46~ & 244~53~16~ & 50~10~52~ & 159~41~35~ & -5~16~27~ \\
Ceres & 11~~01~~14~ & 12~10~36~ & 238~45~52~ & 60~32~34~ & 161~46~26~ & 5~26~43~ \\
Iris & 11~~09~~10~ & -1~14~44~ & 258~58~11~ & 52~37~38~ & 168~47~48~ & -6~10~06~ \\
Amphitrite & 11~~12~~56~ & 17~02~15~ & 233~01~02~ & 65~31~55~ & 162~28~51~ & 11~01~30~ \\
Patientia & 11~~18~~07~ & 9~34~37~ & 248~25~02~ & 62~08~09~ & 166~37~15~ & 4~39~55~ \\
Neptune & 11~~28~~04~ & -7~48~37~ & 271~00~13~ & 49~42~29~ & 175~47~29~ & -10~20~10~ \\
Hebe & 11~~46~~52~ & -0~59~25~ & 272~43~25~ & 57~55~10~ & 177~22~49~ & -2~12~53~ \\
Davida & 11~~55~~49~ & 11~54~11~ & 261~34~52~ & 70~06~40~ & 174~15~10~ & 10~29~35~ \\
Cybele & 12~~00~~21~ & 12~22~41~ & 263~26~31~ & 71~08~58~ & 175~05~28~ & 11~22~38~ \\
Neptune & 12~~00~~43~ & 2~37~33~ & 275~56~20~ & 62~40~04~ & 179~07~12~ & 2~28~50~ \\
Cybele & 12~~03~~06~ & -13~02~02~ & 286~08~37~ & 48~03~43~ & 185~58~03~ & -11~38~03~ \\
Neptune & 12~~11~~14~ & -3~14~42~ & 284~58~44~ & 58~01~51~ & 183~51~56~ & -1~51~41~ \\
Wratislavia & 12~~14~~11~ & -2~55~29~ & 286~10~10~ & 58~32~08~ & 184~24~56~ & -1~16~28~ \\
Fortuna & 12~~20~~58~ & -1~22~22~ & 288~41~02~ & 60~26~51~ & 185~21~23~ & 0~49~22~ \\
Ceres & 12~~21~~50~ & 5~23~20~ & 285~25~25~ & 67~03~35~ & 182~51~45~ & 7~06~52~ \\
Aglaja & 12~~27~~03~ & 2~27~40~ & 290~10~28~ & 64~30~53~ & 185~14~00~ & 4~56~44~ \\
Daphne & 12~~32~~16~ & 6~57~25~ & 291~15~10~ & 69~10~32~ & 184~37~58~ & 9~35~15~ \\
Thalia & 12~~44~~00~ & -3~45~59~ & 300~35~32~ & 58~48~37~ & 191~35~11~ & 0~53~13~ \\
Ceres & 13~~08~~18~ & 2~54~00~ & 314~29~59~ & 65~04~09~ & 194~36~54~ & 9~23~11~ \\
Chicago & 13~~10~~03~ & -16~27~29~ & 310~15~06~ & 45~50~46~ & 202~25~33~ & -8~20~43~ \\
Interamnia & 13~~18~~53~ & -3~03~17~ & 317~28~22~ & 58~42~19~ & 199~21~43~ & 4~53~04~ \\
Cava & 13~~22~~06~ & -10~32~57~ & 316~03~07~ & 51~12~44~ & 202~54~26~ & -1~46~05~ \\
Athamantis & 13~~27~~34~ & 1~27~04~ & 324~13~40~ & 62~27~17~ & 199~41~17~ & 9~52~34~ \\
Ino & 13~~28~~52~ & -3~08~24~ & 322~01~39~ & 57~58~59~ & 201~43~08~ & 5~44~11~ \\
Daphne & 13~~30~~38~ & -11~34~48~ & 318~53~31~ & 49~44~04~ & 205~14~09~ & -1~57~01~ \\
Aglaja & 13~~39~~28~ & -6~15~17~ & 324~47~24~ & 54~12~06~ & 205~19~39~ & 3~48~12~ \\
Melete & 13~~45~~06~ & -19~16~05~ & 320~45~11~ & 41~23~42~ & 211~17~57~ & -7~50~48~ \\
Isis & 14~~12~~49~ & 7~45~23~ & 352~11~01~ & 62~05~02~ & 208~10~36~ & 19~51~25~ \\
Melete & 14~~26~~45~ & -11~17~07~ & 337~45~35~ & 44~38~07~ & 218~02~42~ & 3~04~09~ \\
Fides & 14~~31~~33~ & -13~50~35~ & 337~19~37~ & 41~50~21~ & 219~58~05~ & 1~00~51~ \\
Melpomene & 14~~40~~19~ & -7~48~59~ & 344~26~12~ & 45~39~49~ & 220~08~38~ & 7~24~39~ \\
Bellona & 15~~08~~34~ & -12~02~30~ & 348~15~37~ & 38~02~16~ & 228~04~29~ & 5~22~55~ \\
Uranus & 15~~12~~44~ & -5~09~16~ & 355~21~15~ & 42~25~22~ & 227~11~14~ & 12~17~26~ \\
Palma & 15~~17~~14~ & -17~59~26~ & 345~35~23~ & 32~07~22~ & 231~40~26~ & 0~12~05~ \\
Saturn & 15~~31~~58~ & -15~34~25~ & 350~35~36~ & 31~37~41~ & 234~29~23~ & 3~25~22~ \\
Massalia & 15~~39~~08~ & -17~48~43~ & 350~16~28~ & 28~49~20~ & 236~41~01~ & 1~38~58~ \\
Euphrosyne & 16~~05~~18~ & -11~59~28~ & 0~01~59~ & 28~14~54~ & 241~39~08~ & 8~40~22~ \\
Eleonora & 16~~17~~46~ & -13~55~52~ & 0~33~04~ & 24~40~25~ & 245~01~39~ & 7~19~20~ \\
Dione & 16~~27~~22~ & -19~44~08~ & 357~19~29~ & 19~11~41~ & 248~15~38~ & 1~58~32~ \\
Jupiter & 16~~49~~42~ & -19~10~27~ & 1~12~54~ & 15~25~14~ & 253~24~08~ & 3~15~39~ \\
Europa & 17~~04~~44~ & -16~10~36~ & 5~53~47~ & 14~16~03~ & 256~39~03~ & 6~37~38~ \\
Chaldaea & 17~~09~~43~ & -22~13~01~ & 1~31~08~ & 9~50~44~ & 258~22~35~ & 0~43~09~ \\
Chaldaea & 17~~10~~22~ & -22~13~47~ & 1~35~42~ & 9~42~60~ & 258~31~32~ & 0~43~09~ \\
Carlova & 17~~11~~30~ & -31~19~26~ & 354~16~49~ & 4~13~49~ & 259~33~05~ & -8~19~15~ \\
Flora & 17~~18~~19~ & -18~10~20~ & 6~04~02~ & 10~27~46~ & 260~04~05~ & 4~54~40~ \\
Thisbe & 17~~34~~27~ & -22~35~50~ & 4~24~04~ & 4~53~02~ & 264~06~09~ & 0~42~50~ \\
Massalia & 17~~35~~19~ & -16~02~40~ & 10~06~06~ & 8~11~16~ & 264~01~21~ & 7~16~10~ \\
Carlova & 18~~08~~16~ & -21~32~39~ & 9~18~29~ & -1~18~59~ & 271~55~18~ & 1~53~02~ \\
Papagena & 18~~13~~16~ & 9~03~60~ & 37~03~53~ & 12~02~43~ & 273~52~51~ & 32~27~50~ \\
Berbericia & 18~~18~~32~ & -23~43~42~ & 8~31~17~ & -4~26~29~ & 274~14~23~ & -0~21~17~ \\
Saturn & 19~~21~~02~ & -26~55~24~ & 11~39~58~ & -18~34~54~ & 288~02~46~ & -4~44~31~ \\
Nemausa & 19~~22~~52~ & -20~18~05~ & 18~14~10~ & -16~23~49~ & 289~23~14~ & 1~45~25~ \\
Victoria & 19~~34~~33~ & -22~34~06~ & 17~09~42~ & -19~47~10~ & 291~44~06~ & -0~53~33~ \\
Hygiea & 19~~39~~12~ & -28~07~41~ & 12~00~14~ & -22~46~21~ & 291~51~33~ & -6~32~55~ \\
Vesta & 19~~44~~49~ & -17~09~29~ & 23~27~30~ & -19~53~53~ & 295~01~25~ & 4~02~09~ \\
Saturn & 20~~25~~57~ & -14~11~06~ & 30~46~19~ & -27~47~01~ & 305~21~16~ & 4~53~27~ \\
Aurora & 20~~30~~03~ & -7~57~06~ & 37~34~38~ & -26~02~09~ & 307~51~45~ & 10~41~45~ \\
Papagena & 20~~30~~43~ & -28~19~40~ & 15~43~03~ & -33~38~16~ & 303~01~05~ & -9~06~09~ \\
Davida & 20~~37~~04~ & -16~50~20~ & 29~10~08~ & -31~17~10~ & 307~18~16~ & 1~39~46~ \\
Astraea & 20~~39~~53~ & -8~44~37~ & 38~00~16~ & -28~33~47~ & 310~02~47~ & 9~18~56~ \\
Cava & 20~~42~~24~ & -18~50~03~ & 27~31~44~ & -33~12~05~ & 308~01~21~ & -0~35~26~ \\
Chicago & 20~~50~~21~ & -8~57~51~ & 39~06~38~ & -30~58~29~ & 312~30~38~ & 8~24~57~ \\
Themis & 21~~03~~18~ & -19~31~48~ & 28~51~08~ & -38~03~57~ & 312~34~52~ & -2~36~14~ \\
Io & 21~~17~~58~ & -15~57~27~ & 34~46~19~ & -40~01~20~ & 316~58~14~ & -0~12~59~ \\
Dione & 21~~18~~06~ & -23~41~38~ & 25~04~01~ & -42~38~05~ & 314~38~54~ & -7~35~54~ \\
Iris & 21~~28~~39~ & -15~05~14~ & 37~10~17~ & -42~03~07~ & 319~41~22~ & -0~10~59~ \\
Loreley & 21~~29~~52~ & -26~16~36~ & 22~31~29~ & -45~52~12~ & 316~23~40~ & -10~53~11~ \\
Amphitrite & 21~~37~~23~ & -14~26~59~ & 39~07~56~ & -43~43~18~ & 321~53~28~ & -0~15~02~ \\
Hygiea & 21~~41~~05~ & -20~01~50~ & 32~14~23~ & -46~36~38~ & 320~54~25~ & -5~48~54~ \\
Fortuna & 21~~47~~12~ & 5~46~46~ & 63~04~34~ & -34~58~56~ & 331~05~24~ & 18~00~21~ \\
Alexandra & 21~~52~~02~ & -21~03~50~ & 31~59~37~ & -49~22~08~ & 322~59~27~ & -7~38~47~ \\
Flora & 22~~03~~01~ & -1~42~44~ & 58~38~59~ & -42~43~16~ & 332~11~43~ & 9~36~19~ \\
Davida & 22~~17~~30~ & -0~01~40~ & 63~34~53~ & -44~30~29~ & 336~14~09~ & 9~52~54~ \\
Eleonora & 22~~30~~08~ & -10~17~20~ & 53~51~56~ & -53~09~01~ & 335~24~43~ & -0~49~05~ \\
Germania & 22~~45~~13~ & -14~13~16~ & 51~15~39~ & -58~20~10~ & 337~22~00~ & -5~50~27~ \\
Hebe & 22~~47~~60~ & 10~22~26~ & 81~01~41~ & -42~11~03~ & 347~30~31~ & 16~37~37~ \\
Thalia & 22~~50~~15~ & -8~14~16~ & 61~42~34~ & -56~01~22~ & 340~47~03~ & -0~46~19~ \\
Alexandra & 23~~04~~20~ & -6~47~27~ & 67~52~12~ & -57~44~27~ & 344~33~42~ & -0~46~18~ \\
Harmonia & 23~~18~~57~ & -5~15~21~ & 74~53~45~ & -59~11~27~ & 348~30~36~ & -0~46~20~ \\
Hesperia & 23~~50~~09~ & -14~35~13~ & 73~07~30~ & -71~14~56~ & 351~51~23~ & -12~23~13~ \\
\end{longtable}

\subsection{Earth shine}\label{sec:cavearth}

\begin{figure}[ht]
  \begin{center}
    \FigureFile(80mm,200mm){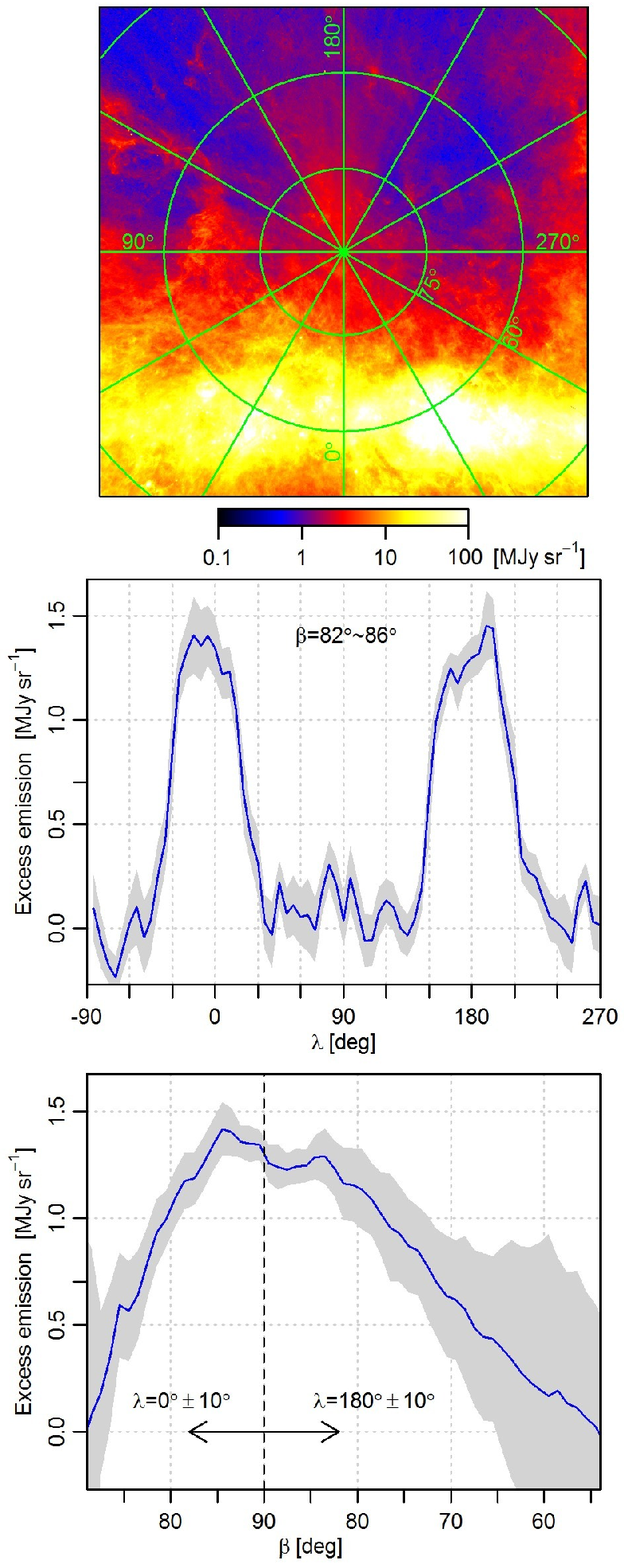}
  \end{center}
  \caption{Top panel -- WIDE-S image of the north ecliptic
    pole. A fan-like distribution of the excess emission due to the earth shine is recognisable.
    Middle and bottom panels -- cross-sections of the excess emission
    in WIDE-S image.
The shaded areas denote the standard deviation of the profile.
  }
  \label{fig:earthshine}
\end{figure}

During the survey observation, the pointing direction of the telescope was kept orthogonal to the
sun-earth direction and away from the centre of the earth, so that we
can minimise the heat input from the earth to the telescope ($\S$\ref{sec:observation}).
However, since the inclination angle of the satellite's orbit was not equal to 90\degree,
the satellite direction was not precisely in the opposite direction to
the centre of the earth and had a small time variation.
This resulted in a small variation in the viewing angle of the earth
limb from the telescope.
The viewing angle became minimum in the observation around the north ecliptic pole in
the summer solstice period, caused the illumination
of the earth's thermal radiation to the top of the telescope
structure.
This thermal radiation was detected as the excess emission in the all-sky
survey image, and is most conspicuous in the WIDE-S band
images with a maximum intensity of $\sim 1.5$ [MJy sr$^{-1}$].
The contaminated regions are spread like a fan around the north ecliptic
pole at around $\lambda$: $-30^{\circ}$ -- $+20^{\circ}$, $\beta
\geq 71^{\circ}$ and $\lambda$: $+150^{\circ}$ -- $+210^{\circ}$, $\beta
\geq 54^{\circ}$.
The cross-section of the spatial profile of the excess emission
is shown in Figure~\ref{fig:earthshine}.
Since the spatial profile is not well determined as the scattering
path of the earth shine in the telescope is not fully studied yet,
we leave the emission in the production image and ask an attention of
the users.
Assuming 300 [K] black-body spectrum for the earth shine, the excess
emission of $1.5$ [MJy sr$^{-1}$] in the WIDE-S band corresponds to
2.2, 0.13, and 0.05 [MJy
sr$^{-1}$] in N60, WIDE-L, and N160 bands, respectively.

\section{Data Release}\label{sec:datarelease}

We have made the first data release of our all-sky survey data to the
public in December 2014. The data are released as $6\degree\times6\degree$ FITS
format intensity image tiles that cover the whole sky,
with supporting data showing the standard deviation
of intensity, data sample numbers, and number of spatial scans in the
same 2-dimensional FITS files.
All the data can be retrieved from the following web site:
http://www.ir.isas.jaxa.jp/AKARI/Observation/

\section{Conclusion}\label{sec:conclusion}

We provide full sky images of the {\it AKARI} FIR survey at 65
\micron, 90 \micron, 140 \micron\ and 160 \micron.
Together with the $> 99$\% coverage of the whole sky, the high spatial
resolution from 1\arcmin\ to 1.5\arcmin\ of the {\it AKARI} FIR survey
reveals the large-scale distribution of ISM with the great detail.
Comprehensive wavelength coverage from 50 \micron\ to 180 \micron\ with four
photometric bands provides SED information at the peak of the dust
continuum emission, enabling us to make precise evaluation of
its temperature, which leads to a detailed investigation of the total
amount of dust particles and its irradiation environment.
The {\it AKARI} FIR images are a new powerful resource from which to investigate
the detailed nature of ISM from small scales to the full sky.

The authors are grateful to the anonymous referee for the useful
discussions and suggestions.
This research is based on observations with
{\it AKARI}, a JAXA project with the participation of
ESA.
This work has been supported by JSPS KAKENHI Grant Number 19204020,
21111005, 25247016, and 25400220.
The authors are grateful to Dr. Sunao Hasegawa for the evaluation of
Planets' and Asteroids' positions.

\end{document}